\documentclass[twocolumn,pra,longbibliography]{revtex4-1}
\usepackage{graphicx}
\usepackage{float}
\usepackage{dcolumn}
\usepackage{bm}
\usepackage{color}
\usepackage{amsmath}
\usepackage{amssymb}
\usepackage{amsfonts}
\usepackage{esint}
\usepackage{times}
\usepackage{xcolor}
\usepackage{phaistos}
\usepackage{braket}
\usepackage{comment}
\usepackage{multirow}

\begin{document}

\title{High-fidelity $\sqrt{i\text{SWAP}}$ gates using a fixed coupler driven by two microwave pulses}

\author{Peng Xu$^{1}$}\email{pengxu@njupt.edu.cn}
\author{Haitao Zhang$^{2,3}$}
\author{Shengjun Wu$^{2,3}$}\email{sjwu@nju.edu.cn}
\affiliation{$^1$ Institute of Quantum Information and Technology, Nanjing University of Posts and Telecommunications, Nanjing 210003, China\\
$^2$ National Laboratory of Solid State Microstructures and School of Physics, Collaborative Innovation Center
of Advanced Microstructures, Nanjing University, Nanjing 210093, China\\
$^3$ Hefei National Laboratory, Hefei 230088, China}


\date{\today}

\begin{abstract}

 Attaining high-fidelity two-qubit gates represents a pivotal quantum operation for the realization of large-scale quantum computation and simulation. In this study, we propose a microwave-control protocol for the implementation of a two-qubit gate employing two transmon qubits coupled via a fixed-frequency transmon coupler. This protocol entails applying two microwave pulses exclusively to the coupler, thereby inducing interaction between the fixed-frequency transmon qubits. This interaction facilitates the realization of $\sqrt{i\text{SWAP}}$ gates. Additionally, we explore the implementation of the gate scheme in two distinct qubit architectures. Demonstrating with experimentally accessible parameters, we show that high-fidelity $\sqrt{i\text{SWAP}}$ gates can be achieved.

\end{abstract}

\maketitle




\section{Introduction}

Due to the intrinsic properties of quantum mechanics, quantum computers can solve complex problems that pose significant challenges for classical computers \cite{ShorPW1994, ShorPW1997, WuYPRL2021, Bermejo-VegaPRX2018, HuangHYScience2022}. Substantial progress has been made in the development of technologies for constructing large-scale quantum computers over the past few decades. Various quantum physical systems have been explored for the implementation of quantum computers, including photons \cite{JLOBrienScience2007, GJPrydeAPR2019}, trapped ions \cite{RBlattPR2008}, superconducting quantum circuits \cite{KrantzPAPR2019, KjaergaardM2019, AlexandreBlaisPRA2004, AlexandreBlaisPRA2007, JClarkeNature2008, ABlaisRMP2021}, and semiconductors \cite{JRPettaRMP2023}. Among these physical systems, superconducting quantum circuits based on circuit quantum electrodynamics \cite{ABlaisRMP2021} have emerged as promising platforms for scalable quantum information processors, with several significant experimental milestones achieved over recent decades \cite{MASillanpNature2007, JMajerNature2007, AONiskanenScience2007, PlantenbergNature2007, LDiCarloNature2009, MAnsmannNature2009, FAruteNature2019}. Notably, superconducting quantum circuits offer advantages such as the use of solid-state qubits, allowing for tailored design and fabrication, as well as the potential for constructing qubit arrays in two-dimensional geometries \cite{MGongScience2021} or employing three-dimensional combinations \cite{DRWYostnpjQI2020, JLMallekarXiv2103, SKosenQST2022}, thereby facilitating the efficient realization of scalable quantum processor architectures.

In recent years, there has been a rapid increase in the number of superconducting qubits, accompanied by enhancements in their quality. A significant experimental milestone, referred to as quantum supremacy \cite{BoixoSNP2018}, was initially achieved using superconducting quantum circuits in 2019 \cite{FAruteNature2019}. As a result of the swift advancement of quantum technologies, we have transitioned into the era of noisy intermediate-scale quantum (NISQ) computing \cite{PreskillJQauntum2018}. The ultimate objective is to realize a universal quantum computer capable of executing various complex computations.

It is well established that two-qubit gates, combined with a single-qubit quantum gate, is sufficient for universal quantum computation \cite{DDeutschPNAS1995, ABarencoPNAS1995, DPDiVincenzoPRA10151995, TSleatorPRL1995, SLloyd1995, ABarencoPRA34571995}. However, achieving high-fidelity two-qubit gates remains notoriously challenging in superconducting qubit systems \cite{XGuPhysRep2017}. For superconducting qubits, one primary approach to implementing two-qubit gates involves dynamically tuning the qubit frequency \cite{JMajerNature2007, LDiCarloNature2009, LDiCarloNature2010, RBarendsNature2014}, resonator frequency \cite{DCMcKayPRAppl2016, MRothPRA2017}, or the qubit-qubit coupling strength \cite{YChenPRL2014, FeiYanPRAp2018}. However, the frequency-tunable method requires additional control lines, potentially adding additional hardware overhead and inducing a decoherence channel from operating off the qubit's optimal point. Another method entails utilizing microwave pulses to select and control qubit transitions to implement two-qubit gates \cite{JMChowNJP2013, EBarnesPRB2017, SPPremaratnePRA2019, CRigettiPRB2010, JMChowPRL2011, JLAllenPRA2017, NakamuraPRL2606012023}. Here, both single- and two-qubit gates can be realized for fixed-frequency qubits and fixed qubit-qubit couplings, giving rise to the qubit architecture with a longer coherence time. Nevertheless, residual couplings and spectral crowding problems often lead to lengthy gate times required to achieve high-fidelity microwave-activated gates. This suggests that a trade-off exists between gate speed and gate fidelity. Thus, realizing fast, high-fidelity microwave-activated two-qubit gates is still challenging in scalable qubit systems with fixed-frequency qubits and fixed couplings.

In this paper, we investigate the implementation of a microwave-activated two-qubit gate in superconducting quantum circuits by applying two microwave pulses solely to drive the coupler qubit. Through the application of these pulses on the coupler qubit, an effective interaction between the two qubits can be achieved. Given fixed system parameters, such as qubit frequencies and qubit-coupler coupling strength, achieving high fidelity in the two-qubit $\sqrt{i\text{SWAP}}$ gate relies on tailoring the pulse shapes and controlling microwave pulse parameters. This approach, compared to frequency-tunable methods, minimizes circuit complexity without requiring additional flux control lines, thereby reducing decoherence stemming from flux noise. Moreover, the fixed coupling strength between transmon qubits and the coupler, established through simple capacitances, enhances the reliability of the superconducting quantum circuit. The resultant effective interaction facilitates the swapping of excitations between the two qubits, rendering the $\sqrt{i\text{SWAP}}$ gate \cite{JianxinChenPRL130070601} expressible as,
\begin{equation}
{\rm \sqrt{\textit{i}\text{SWAP}}} \equiv  
\left[
\begin{array}{cccc}  
1 & 0 & 0 & 0 \\
0 & 1/\sqrt{2} & -i/\sqrt{2} & 0 \\
0 & -i/\sqrt{2} & 1/\sqrt{2} & 0 \\
0 & 0 & 0 & 1
\end{array}
\right ].                              
\end{equation}

We examine the two-qubit gate within two distinct qubit architectures designated as ABA and ABC. In this context, A and C represent the transmon qubits, with B denoting the transmon coupler. The ABA-type architecture indicates comparable frequencies between the qubits, while the frequency of the coupler resides in a separate frequency band. Conversely, the ABC-type architecture entails three distinct frequencies for the qubits and coupler, with the coupler's frequency positioned between the frequencies of the two qubits. For the ABA-type architecture, we show that the two-qubit $\sqrt{i\text{SWAP}}$-like gate with a fidelity exceeding 99.92\% can be achieved \cite{BarendsPRL2105012019, BFoxenPRL125120504, MGVavilovPRXQ2020345, MJSPRA108052619}. Similarly, for the ABC-type architecture, the average gate fidelity can also reach 99.93\%. The proposed gate scheme readily extends to the scenario in which multi-qubits are coupled via a single fixed-frequency coupler, enabling the realization of $\sqrt{i\text{SWAP}}$ gates for arbitrary pairs of qubits. Furthermore, by increasing the driving strength of microwave pulses or optimizing system parameters, we can shorten the gate time.

The paper is organized as follows. In Sec. II, we introduce the system Hamiltonian for the superconducting quantum circuit, which comprises two transmon qubits commonly coupled to a transmon coupler and driven by two microwave fields. Subsequently, based on the perturbation theory, we give an analytic expression of the effective two-qubit interaction. In Sec. III, we discuss the implementation of the $\sqrt{i\text{SWAP}}$ gates for both ABA- and ABC-type architectures. Sec. IV focuses on analyzing the influence of system relaxation and ZZ coupling interaction. Finally, our conclusions are presented in Sec. V. Furthermore, Appendix A presents all evolution paths potentially contributing to the effective coupling strength between qubits. Appendix B offers an analysis of fidelity and leakage for the microwave-driven two-qubit system. Additionally, Appendix C delves into the exploration of the proposed two-qubit gate scheme within multi-qubit systems.

\begin{figure}
\begin{center}
\includegraphics[width=4.5cm, height=4.50cm]{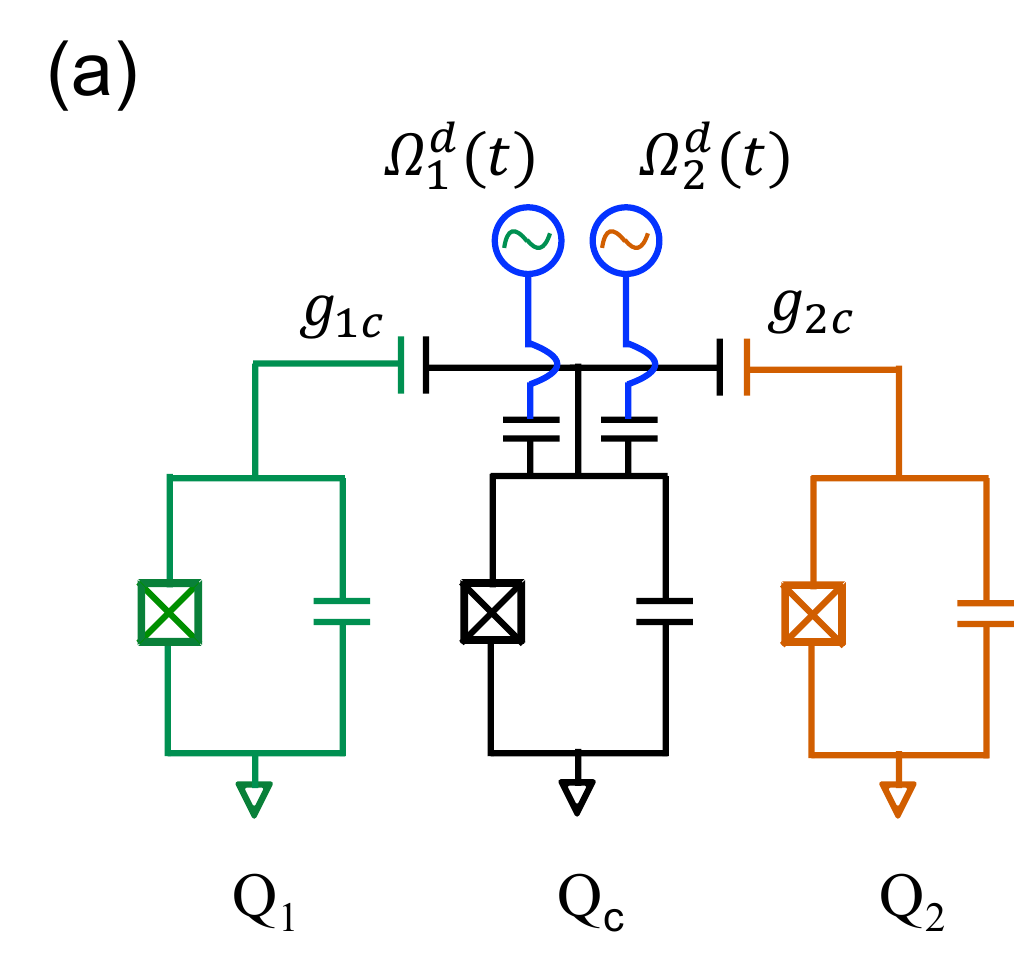}\includegraphics[width=4.50cm, height=4.50cm]{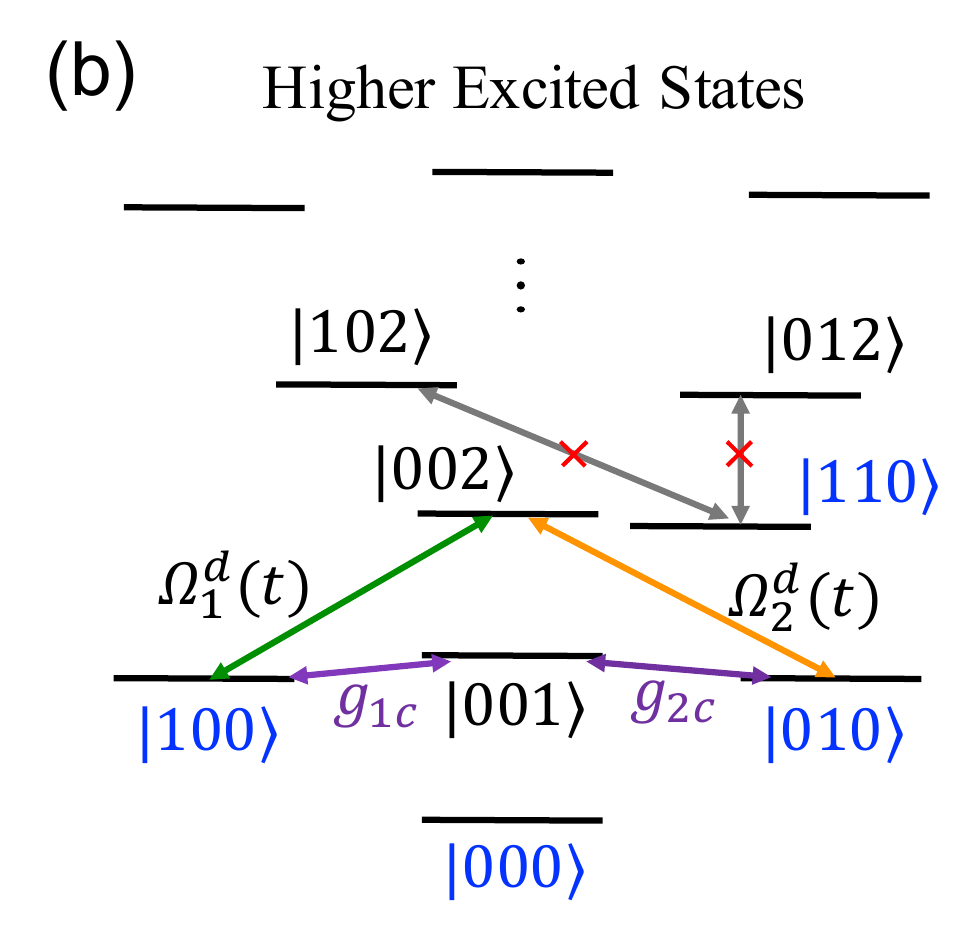}
\end{center}
\caption{Schematic circuit diagram of the proposed pulse-driven two-qubit gate scheme. (a) The superconducting quantum circuit comprises two transmon qubits ($Q_{1,2}$) capacitively coupled to a common coupler qubit $Q_c$, which is driven by two external microwave pulses. (b) A schematic level diagram illustrates the fixed qubit-coupler couplings $g_{kc}$ and microwave drives $\Omega^d_{1(2)}(t)$, facilitating the necessary effective interaction between the target states $|100\rangle$ and $|010\rangle$. The interaction between $|102\rangle$ (or $|012\rangle$) and $|110\rangle$ is unwanted. 
\label{FigOne}}
\end{figure}

\section{Superconducting Transmon Circuits: A Study of Hamiltonian Dynamics}

In superconducting circuits, one qubit of particular interest is the transmon \cite{JKochPRA0423192007}, which is derived from the charge-based qubit and exhibits reduced susceptibility to charge noise, leading to longer coherence times. Our focus lies on a superconducting quantum circuit comprising two superconducting transmon qubits coupled to a transmon coupler qubit, as depicted in Fig. 1(a). To implement the $\sqrt{i\text{SWAP}}$ gate, two microwave pulses are applied to the transmon coupler. Considering the anharmonicity, the transmon qubits and coupler can be represented as weakly anharmonic oscillators. The Hamiltonian governing the interaction of two transmons coupled to a coupler can be expressed as (hereafter $\hbar = 1$)
\begin{equation}
\begin{aligned}
H_s = &\sum_{j =1, 2, c}  (w_j q^\dagger_j q_j + \frac{\alpha_j}{2} q^\dagger_j q^\dagger_j q_j q_j ) \\+ &\sum_{k=1,2}g_{kc}(q^\dagger_k + q_k)(q^\dagger_c + q_c),
\end{aligned}
\end{equation} 
where $\omega_j$ and $\alpha_j$ denote the frequency and anharmonicity of the transmon qubit $Q_j$, respectively. The coupling between transmon $Q_k$ and the coupler $Q_c$ is denoted by $g_{kc}$. The operators $q_j$ ($q_j^{\dagger}$) correspond to the annihilation (creation) operators associated with qubit $Q_j$. To account for the impact of higher levels, each qubit and coupler are truncated to four levels, namely, $|0\rangle$, $|1\rangle$, $|2\rangle$, and $|3\rangle$. Throughout the paper, we discuss the complete system state $|Q_1, Q_2, Q_c\rangle$, reflecting the states of qubits 1 and 2, as well as the coupler qubit $Q_c$. When focusing solely on the subspace defined by the transmon qubits, we adopt the simplified notation $|Q_1, Q_2\rangle$ for brevity.

To achieve $\sqrt{i\text{SWAP}}$ gate, additional control over the quantum system is necessary to induce the requisite interactions. In this context, we explore the application of two microwave pulses to drive the coupler qubit, thereby leading to an effective interaction between $Q_1$ and $Q_2$. The driving Hamiltonian is expressed as follows,
\begin{equation}
\begin{aligned}
H_d =  \sum_{k=1,2}\Omega^d_{k}(t)\text{cos}(\omega^d_{k} t + \phi_k)(q^\dagger_c + q_c). 
\end{aligned}
\end{equation}
Here, $\Omega^d_{k}(t) = \Omega_{k}(0)G^d_k(t)$ represents the time-dependent pulse envelope with driving amplitude $\Omega_{k}(0)$ and pulse shape $G^d_k(t)$, $\omega^d_{k}$ denotes the drive frequency, and the initial phase $\phi_k$ is set to 0 for simplicity. In the context of the driving pulses applied to the coupler, we provide an overview of the system's energy levels, as depicted in Fig. 1(b). Performing two drives on the coupler could enhance the effective interaction between the two qubits and also induce unwanted interactions between the defined computational basis states and non-computational basis states, such as $|102\rangle$, $|012\rangle$, and $|110\rangle$. Utilizing the defined basis states {$|00\rangle$, $|01\rangle$, $|10\rangle$, $|11\rangle$}, the two-qubit swap interaction (with basis states defined in target qubits) can be expressed as,
\begin{equation}
{\rm U_{swap}} \equiv 
\left[
\begin{array}{cccc}  
1 & 0 & 0 & 0 \\
0 & \cos\theta & -i\sin\theta & 0 \\
0 & -i\sin\theta & \cos\theta & 0 \\
0 & 0 & 0 & 1
\end{array}
\right ],                               
\end{equation}
with swap angle $\theta = \pi/4$ transforms into the $\sqrt{i\text{SWAP}}$ gate as depicted in Eq. (1).

\begin{figure}
\begin{center}
\includegraphics[width=4.5cm, height=4.5cm]{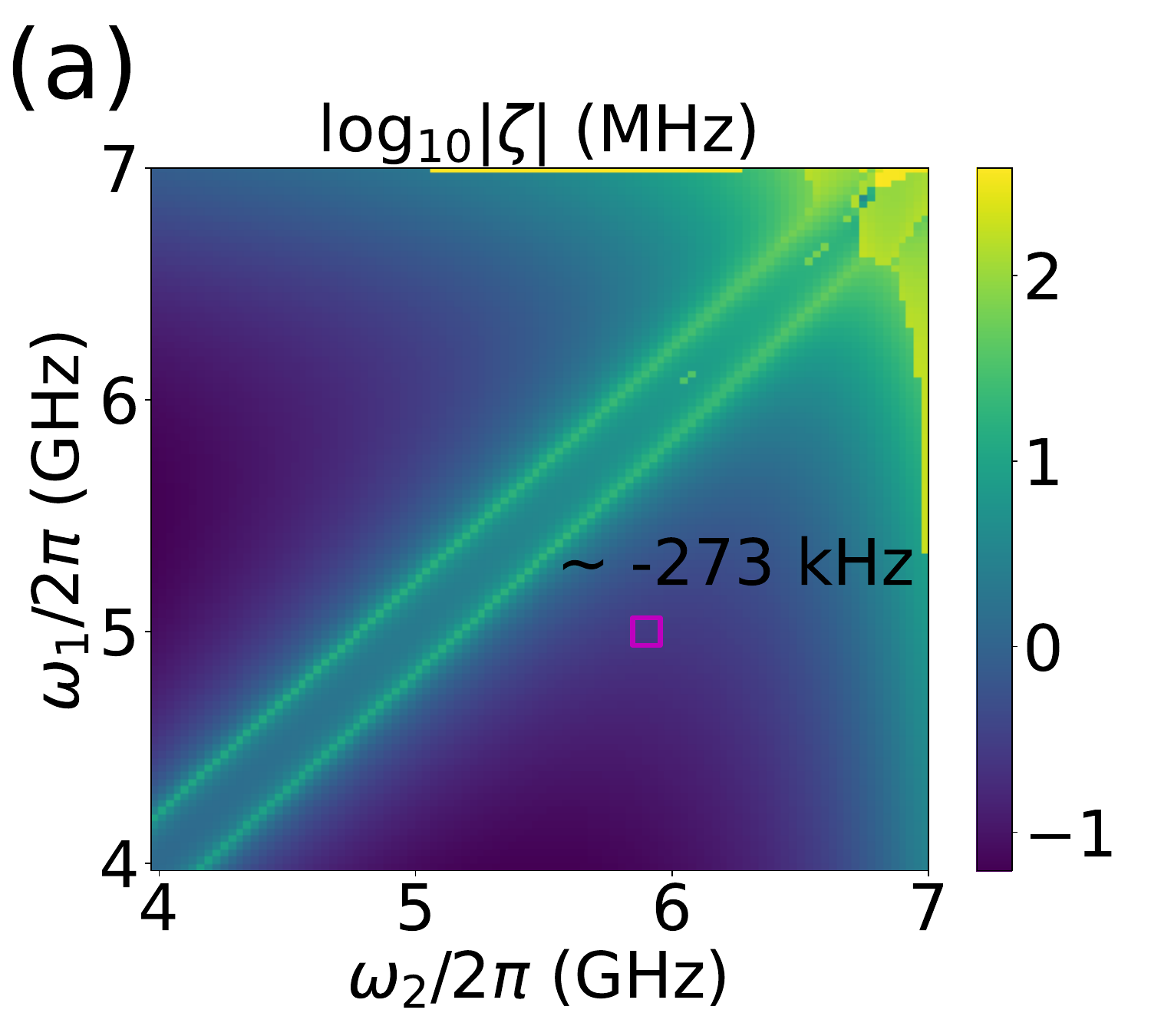}\includegraphics[width=4.5cm, height=4.5cm]{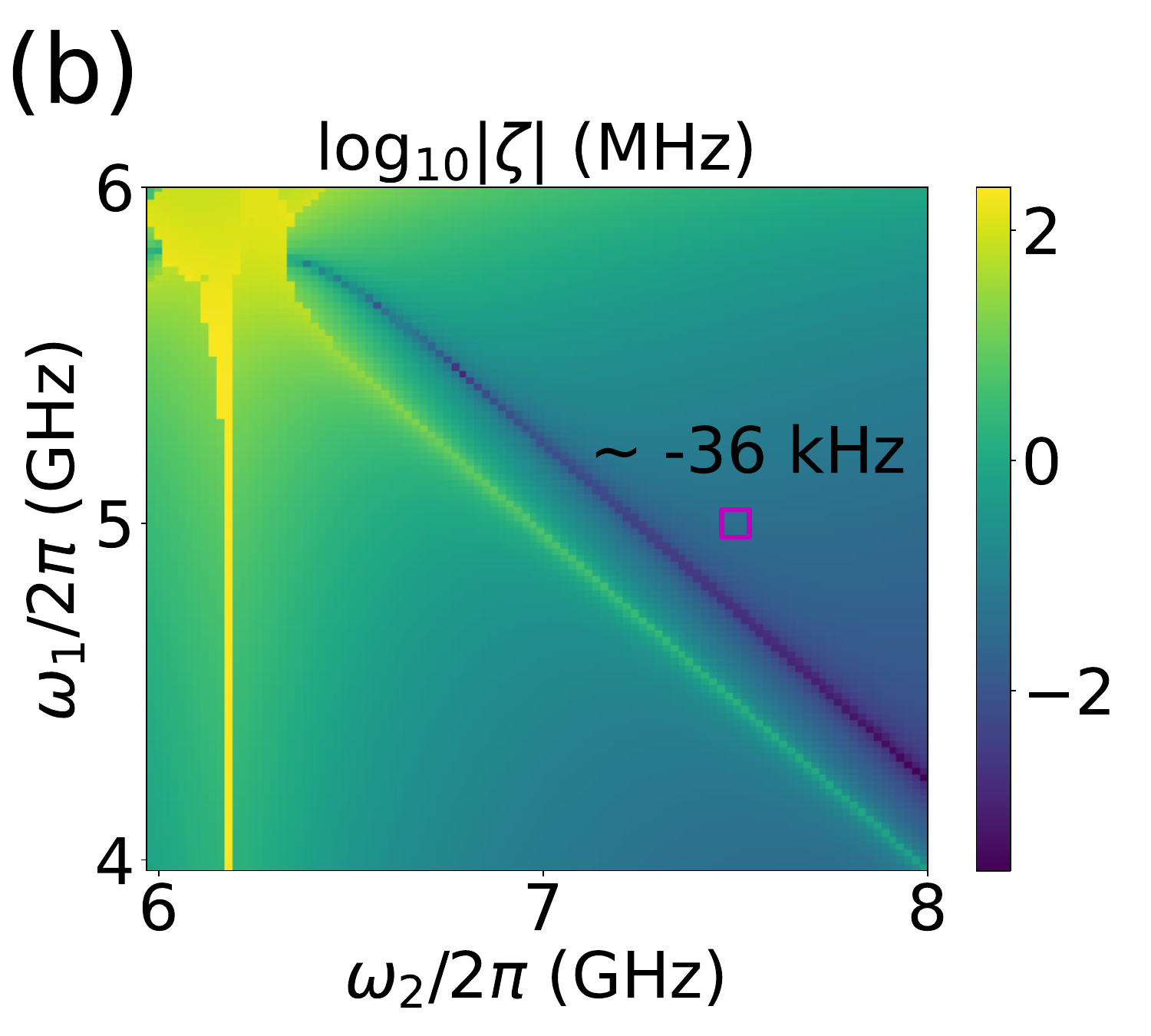}
\end{center}
\caption{The ZZ coupling strength vs the qubits' frequencies $\omega_1$ and $\omega_2$. (a) Illustrates the change of $\text{log}{10}|\zeta|$ for ABA-type architecture with specific system parameters: $\omega_c/2\pi = 7.0$ GHz, $\alpha_{1(2)}/2\pi = -200$ MHz, $\alpha_{c}/2\pi = -400$ MHz, and coupling strengths $g_{1c}/2\pi = 190$ MHz, $g_{2c}/2\pi = 100$ MHz. In this case, the ZZ coupling strength can be suppressed below 100 kHz. (b) Shows the relationship between the ZZ coupling strength and qubits' frequencies $\omega_1$ and $\omega_2$ in the ABC-type architecture, with system parameters $\omega_c/2\pi = 6.2$ GHz, $\alpha_{1(2)}/2\pi = -300$ MHz, $\alpha_{c}/2\pi = -400$ MHz, and coupling strengths $g_{1c}/2\pi = 110$ MHz, $g_{2c}/2\pi = 120$ MHz. Here, the ZZ coupling can be restrained below 10 kHz. The magenta square highlighted in the figures indicates the selected frequencies of the two qubits necessary for realizing the two-qubit $\sqrt{i\text{SWAP}}$ gate. 
\label{FigTwo}}
\end{figure}

It is well known that the $\sqrt{i\text{SWAP}}$ gate relies on XY interactions. Achieving high-fidelity $\sqrt{i\text{SWAP}}$ gate operations necessitates minimizing the ZZ coupling component. The strength of ZZ coupling can be computed using the energy of the first excited state $|\widetilde{mn}\rangle$ ($m, n =$ {0,1}). Here, $|\widetilde{mn}\rangle$ represents the eigenstate of the Hamiltonian $H_s$ with the highest overlap with the bare state $|mn0\rangle$, and its corresponding eigenenergy is denoted as $E_{\widetilde{mn}}$. The ZZ coupling strength can be determined by the following expression
\begin{equation}
\begin{aligned}
\zeta = (E_{\widetilde{11}} - E_{\widetilde{01}} )- (E_{\widetilde{10}} - E_{\widetilde{00}}).
\end{aligned}
\end{equation}
We proceed to examine the variation in ZZ coupling strength concerning the frequencies of the transmon qubits $\omega_1$ and $\omega_2$, as illustrated in Fig. 2. It is observed that the ZZ coupling strength can be significantly mitigated by appropriately selecting the values of qubits frequencies.

To implement the $\sqrt{i\text{SWAP}}$ gate, the frequencies of the driving pulses must meet certain conditions: $\omega^d_1 \approx 2\omega_c + \alpha_c - \omega_1$ and $\omega^d_2 \approx 2\omega_c + \alpha_c - \omega_2$. These drives applied to the coupler induce resonant interactions between the states $|100\rangle \leftrightarrow |002\rangle$ and $|010\rangle \leftrightarrow |002\rangle$, respectively. By utilizing the intermediate state $|002\rangle$, effective interactions between the states $|100\rangle$ and $|010\rangle$ can be achieved. However, undesired populations in state $|002\rangle$ resulting from the qubit-coupler interactions need to be avoided. To mitigate resonant interactions between $Q_{1(2)}$ and $Q_{c}$, a detuning $\delta$ between them must be considered. Consequently, the frequencies for the two driving pulses become
\begin{equation}
\begin{aligned}
&\omega^d_1 \approx 2\omega_c + \alpha_c - \omega_1 + \delta,\\
&\omega^d_2 \approx 2\omega_c + \alpha_c - \omega_2 + \delta.
\end{aligned}
\end{equation}

\begin{figure}
\begin{center}
\includegraphics[width=4.4cm, height=4.3cm]{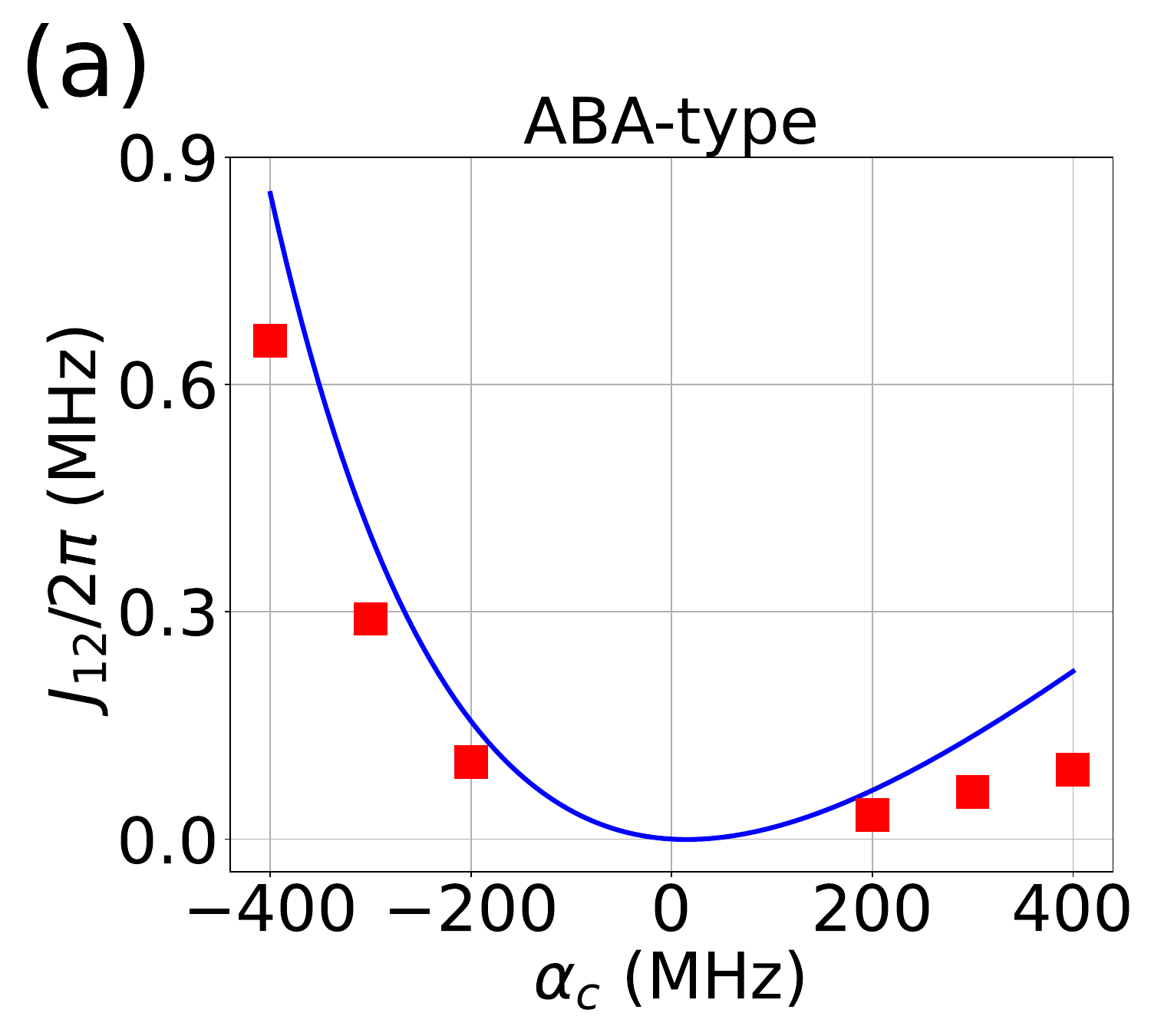}\includegraphics[width=4.4cm, height=4.3cm]{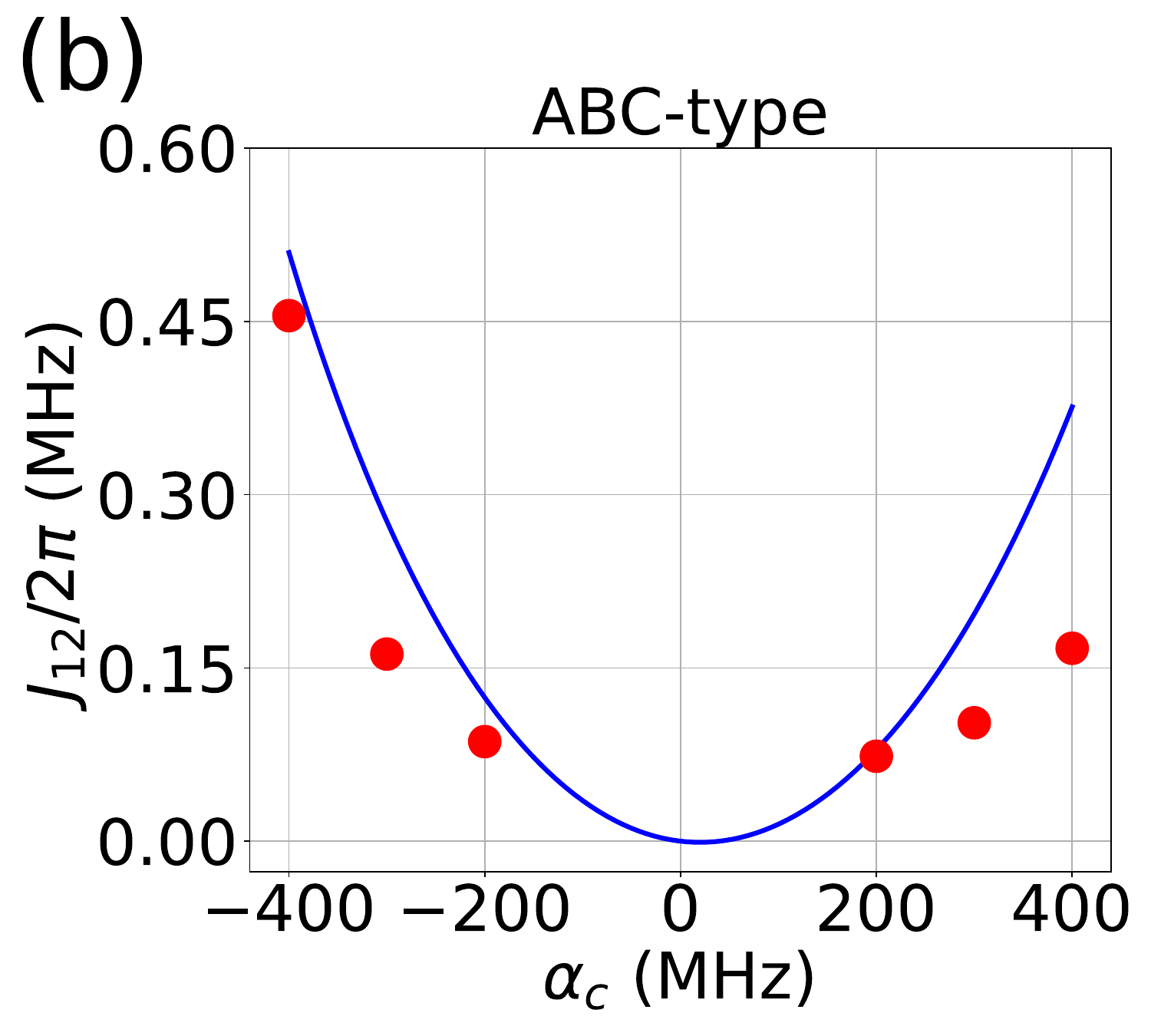}
\end{center}
\caption{The effective coupling strength as a function of the coupler anharmonicity $\alpha_c$. (a) Demonstrates the variation of $J_{12}$ for the ABA-type architecture, employing the system parameters detailed in Table I and pulse parameters $\Omega_{1(2)}(0)/2\pi$ = 160 MHz, $\delta/2\pi = -30$ MHz. The red square points denote the effective coupling strength stemming from the system dynamics. (b) Depicts the effective coupling strength against the coupler anharmonicity $\alpha_c$ in the ABC-type architecture, utilizing the system parameters outlined in Table II and pulse parameters $\Omega_{1(2)}(0)/2\pi$ = 160 MHz, $\delta/2\pi = -40$ MHz. The red dots signify the effective coupling strength arising from the system dynamics. The anharmonicity of the coupler qubit for both ABA and ABC architectures is set to $-400$ MHz to facilitate the implementation of the two-qubit $\sqrt{i\text{SWAP}}$ gate. 
\label{FigThree}}
\end{figure}

The effective interaction strength between $|100\rangle$ and $|010\rangle$ is facilitated through two-photon transitions \cite{KlausAndersPRA0223112000} involving five intermediate states: $|002\rangle$, $|110\rangle$, $|100\rangle$, $|010\rangle$, and $|000\rangle$. Assuming the system operates in the dispersive regime where the detuning between the qubit-coupler and coupler-pulse frequencies is significantly larger than the coupling and driving strengths, i.e., $\omega_{1(2)} - \omega_{c} \gg g_{1(2)c}$ and $\omega_{c} - \omega^d_{1(2)} \gg \Omega_{1(2)}(0)$. Accordingly, the aforementioned transition paths result in effective coupling interactions between states $|100\rangle$ and $|010\rangle$ with a strength denoted as $J_{12}$ \cite{AFKockumPRA0638492017}
\begin{equation}
\begin{aligned}
J_{12} = &\frac{g_1\sqrt{2}\Omega_1g_2\sqrt{2}\Omega_2}{4\Delta_1\delta(\Delta_2-\alpha_c-\delta)} +\frac{g_1\sqrt{2}\Omega_12\Omega_2g_2}{-4\Delta_1\delta\Delta_2}&\\
+&\frac{\Omega_1\sqrt{2}g_1\sqrt{2}g_2\sqrt{2}\Omega_2}{-4(\Delta_1-\alpha_c-\delta)\delta(\Delta_2-\alpha_2-\delta)}&\\
+&\frac{\Omega_1\sqrt{2}g_12\Omega_2g_2}{4(\Delta_1-\alpha_c-\delta)\delta\Delta_2}&\\
+&\frac{\Omega_1g_2g_1\sqrt{2}\Omega_2}{4(\Delta_1-\alpha_c-\delta)(\Delta_1+\Delta_2-\alpha_c-\delta)(\Delta_2-\alpha_c-\delta)}&\\
+&\frac{\Omega_1\sqrt{2}\Omega_2g_1g_2}{-4(\Delta_1-\alpha_c-\delta)(\omega_1-\omega_2)\Delta_2}&\\
+&\frac{g_1g_2\Omega_1\sqrt{2}\Omega_2}{-4\Delta_1(\omega_2-\omega_1)(\Delta_2-\alpha_c-\delta)}&\\
+&\frac{g_1\sqrt{2}\Omega_2\Omega_1g_2}{4\Delta_1(-\Delta_1-\Delta_2+\alpha_c+\delta)\Delta_2}&
\end{aligned}
\end{equation}
where $\Delta_{k}=\omega_k - \omega_c$ represents the detuning between qubit-$k$ and the coupler, while $\omega^d_{1,2}$ denotes the frequency of the driving mode applied to the coupler.

To achieve a high-fidelity $\sqrt{i\text{SWAP}}$ gate in the ABA-type qubits architecture, we adopt the system parameter values outlined in Table I (Sec. III A). Concurrently, we aim to suppress the ZZ coupling strength to approximately $-273$ kHz. In the case of the ABC-type qubits framework, we select the system parameter values provided in Table II (Sec. III B), with the target of suppressing the residual ZZ coupling strength to about $-36$ kHz. For the ABA-type architecture, the parameters for the drive pulses applied to the coupler qubit are configured as $\Omega_{1(2)}(0)/2\pi$ = 160 MHz and $\delta/2\pi = -30$ MHz. Similarly, for the ABC-type framework, the pulse parameters are set to $\Omega_{1(2)}(0)/2\pi$ = 160 MHz and $\delta/2\pi = -40$ MHz. With the system and pulse parameters determined, we analyze the effective interaction strength $J_{12}$ between states $|100\rangle$ and $|010\rangle$ as a function of the coupler anharmonicity, as illustrated in Fig. 3. One can observe that the greater the anharmonicity of the coupler, the stronger the effective interaction becomes. When the anharmonicity is set to 0, there is almost no interaction between $|100\rangle$ and $|010\rangle$.

Given the system and pulse parameters, the effective interaction strength can be obtained through the numerical simulation of the system dynamics, as depicted by the red points in Fig. 3. Notably, the analytical coupling strength obtained from perturbation theory aligns closely with the numerical results. Consequently, the effective coupling strength derived via perturbation theory serves as a reliable estimate for determining the gate time. Particularly, the effective coupling strength $J_{12}$ increases with the rise in the coupler's anharmonicity. We opt to examine the implementation of $\sqrt{i\text{SWAP}}$ gates at the point with an anharmonicity of $-400$ MHz.

\section{Realizing the $\sqrt{i\text{SWAP}}$ Gate in Two-Qubit Energy Architectures}

In the following section, we analyze the implementation of the two-qubit $\sqrt{i\text{SWAP}}$ gate for two distinct types of qubit energy architectures. As previously discussed, achieving the two-qubit $\sqrt{i\text{SWAP}}$ gate necessitates establishing an effective coupling between states $|10\rangle$ and $|01\rangle$, followed by meticulous control of the system dynamics through tailored microwave pulses applied to the coupler. Furthermore, the presence of high energy levels leads to the phenomenon where external impulse drives cause leakage of quantum states from the computational basis to the non-computational basis states. This leakage can significantly impact the fidelity of quantum gate operations. To mitigate this effect, one approach is to apply microwave drives with smooth pulses shape. For the sake of simplicity, we employ a time-dependent Gaussian pulse waveform, which is expressed as follows,
\begin{equation}
\begin{aligned}   
G^d_k(t) =
 \begin{cases}
 \frac{e^{-(t-t_r)^2/(2\gamma^2)} - e^{-t_r^2/(2\gamma^2)}}{1 - e^{-t_r^2/(2 \gamma^2)}}, & t \le t_r \\
 1, & t_r < t < t_p - t_r \\
  \frac{e^{-(t-(t_p - t_r))^2/(2\gamma^2)} - e^{-t_r^2/(2\gamma^2)}}{1 - e^{-t_r^2/(2 \gamma^2)}}, & t_p - t_r \le t \le t_p \\
 0, & \text{others}
 \end{cases}
\end{aligned}
\end{equation}
where $t_r$ represents the ramp time to reach the working point of the microwave pulse, $\gamma$ denotes the variance of the Gaussian pulse, and $t_p$ signifies the plateau time. Taking into account the Gaussian pulse shape, the time-dependent pulse envelope becomes $\Omega^d_k(t) = \Omega_k(0)G^d_k(t)$.

In the dressed state picture, the qubit systems will traverse numerous avoid-crossings caused by undesired microwave-activated interaction during the gate operation. Hence, it is imperative to meticulously design and select appropriate pulse shape parameters to circumvent non-negligible leakage from these avoid-crossings. The precise pulse shape parameters can be chosen through dynamic evolution.

\begin{figure}
\begin{center}
\includegraphics[width=4.60cm, height=4.05cm]{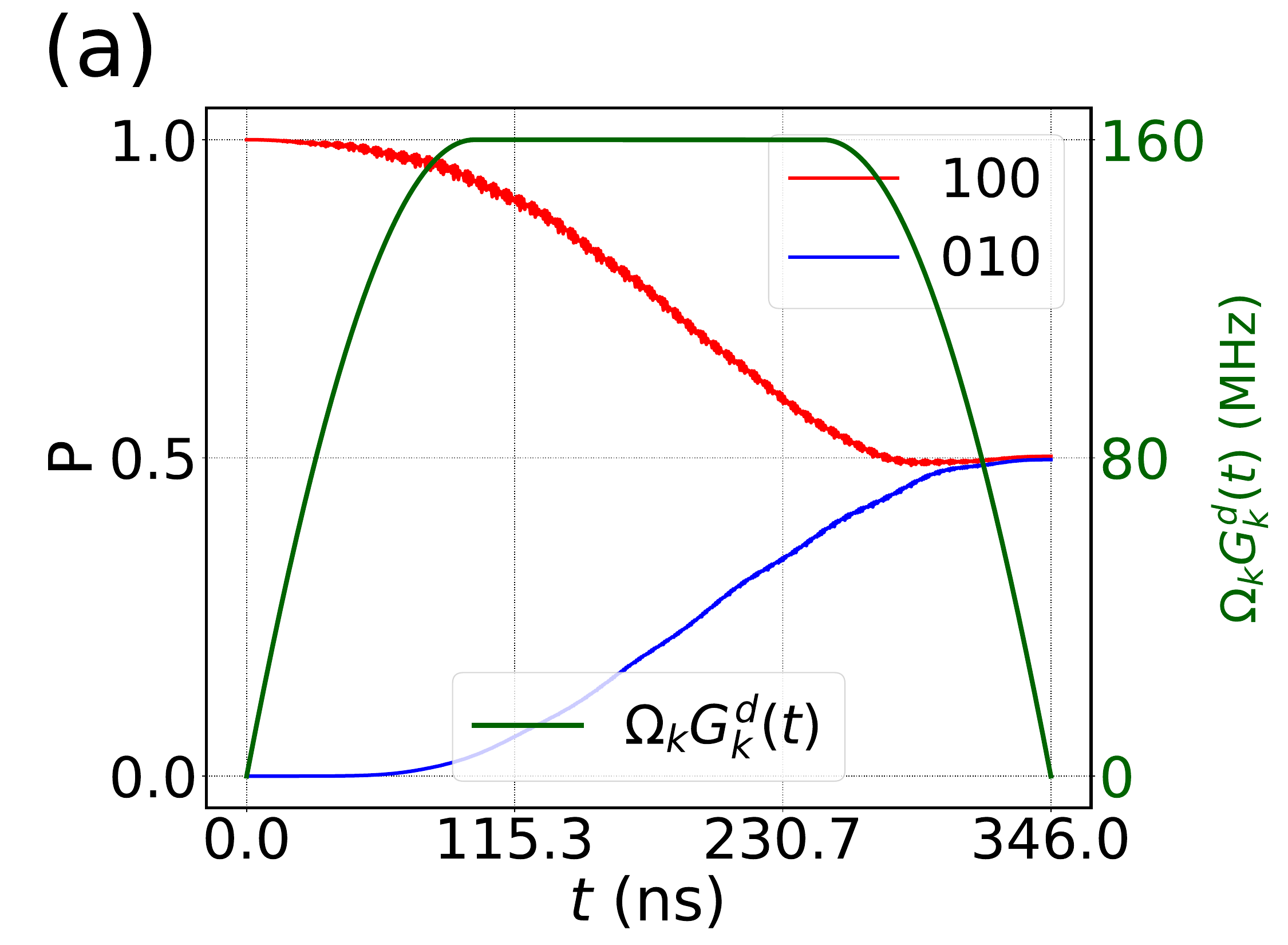}\includegraphics[width=4.0cm, height=4.05cm]{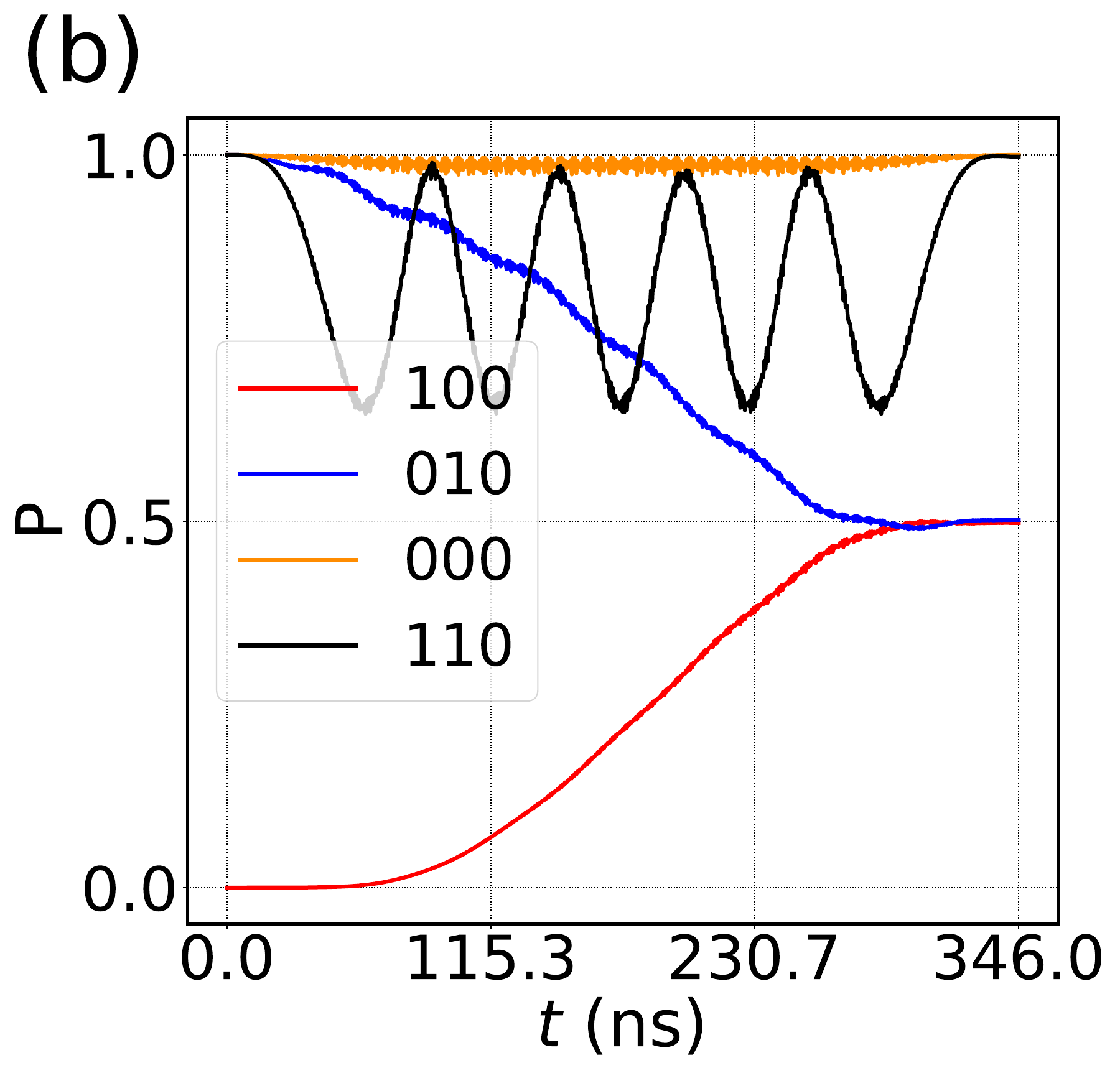}
\end{center}
\caption{The system dynamics during the gate operation. In (a), the change in populations for states $|100\rangle$ and $|010\rangle$ is depicted with the system's initial state set to $|100\rangle$. The solid green line represents the applied time-dependent envelope with $\gamma=196$ ns and $t_{p} = 346$ ns. In (b), the populations of states $|100\rangle$, $|010\rangle$, $|000\rangle$, and $|110\rangle$ are illustrated corresponding to the initial states $|010\rangle$, $|000\rangle$, and $|110\rangle$, respectively. The pulse parameters are configured as $\Omega_{1(2)}(0)/2\pi$ = 160 MHz, $\omega^d_1/2\pi$ $\approx$ 8.654 GHz, and $\omega^d_2/2\pi$ $\approx$ 7.744 GHz, with a fixed relation $t_r = 0.5\gamma$. The values of the system parameters are presented in Table I.
\label{FigFour}}
\end{figure}

\subsection{ABA Architecture: Analysis and Implementation}

\begin{table}[h]
\centering
\caption{System parameters for two-qubit $\sqrt{i\text{SWAP}}$ gate in ABA-type architecture.}
\label{my-label}
\begin{tabular}{@{}lccccc@{}}
\toprule
              & Bare frequency (GHz) & Anharmonicity (MHz)   & Coupling (MHz)  \\
\hline
$Q_1$ &  $\omega_1/2\pi$ = 5.0        &          $\alpha_1/2\pi = -200$    & \multicolumn{1}{c}{\multirow{2}{*}{$g_{1c}/2\pi$ = 190}}    \\
$Q_c$     & $\omega_c/2\pi$ =  7.0        &             $\alpha_c/2\pi = -400$      &   \multicolumn{1}{c}{\multirow{2}{*}{$g_{2c}/2\pi$ = 100}}  \\
$Q_2$ &  $\omega_2/2\pi$ = 5.9        &          $\alpha_2/2\pi = -200$    &                               \\  
\toprule
\end{tabular}
\end{table}

Assuming fixed system parameters upon fabrication of the superconducting transmon qubits, driving applied to the coupler can induce interactions among states $|100\rangle$, $|010\rangle$, and $|002\rangle$. However, resonance between $|100\rangle \leftrightarrow |002\rangle$ and $|010\rangle \leftrightarrow |002\rangle$ is unnecessary, while an effective interaction between $|100\rangle$ and $|010\rangle$ is desired. Hence, a detuning term $\delta$ is essential, as described in Eq. (6). Utilizing the fixed system parameters from Table I and a specified detuning $\delta/2\pi = -30$ MHz, approximate driving frequencies $\omega^d_1/2\pi \approx 8.654$ GHz and $\omega^d_2/2\pi \approx 7.742$ GHz can be derived. However, due to the AC stack effects, these frequencies may shift. Achieving a high-fidelity two-qubit $\sqrt{i\text{SWAP}}$ gate operation necessitates pinpointing the precise driving frequencies. For simplicity, constant-amplitude driving pulses $\Omega_{1(2)}(0) = 160$ MHz are considered in Eq. (3). As illustrated in Eq. (1), the $\sqrt{i\text{SWAP}}$ gate can maximally entangle the system initialized in state $|100\rangle$ (or $|010\rangle$). Initializing the quantum system in state $|100\rangle$, we numerically compute the populations of states $|100\rangle$ and $|010\rangle$ as functions of time $t$ and driving frequency $\omega^d_2$. Through scanning $\omega^d_2$ during system dynamics based on the approximate drive frequencies $\omega^d_{2}$ and system behavior, one precise value of $\omega^d_2$ facilitating maximal entanglement between $Q_1$ and $Q_2$ can be determined. To address this, we examine system dynamics with $\omega^d_2/2\pi \approx 7.744$ GHz for various initial states $|100\rangle$ and $|010\rangle$. While this frequency choice enables the entanglement state between $|100\rangle$ and $|010\rangle$, stray interactions with non-computational basis states may compromise gate operation fidelity between $Q_1$ and $Q_2$.


To minimize gate operation errors, as discussed earlier, it's crucial to design a pulse shape that mitigates the stray interactions between computational basis and non-computational states during gate operation. Notably, there exist subtle relations among the pulse shape parameters $t_r$, $\gamma$, and $t_{p}$ in Eq. (9). For simplicity, we maintain a fixed relation between the ramp time parameter and the variance, setting $t_r = 0.5\gamma$. According to numerical simulations, precise values of pulse shape parameters can be determined, $\gamma=196$ ns and $t_{p} = 346$ ns. These parameter values are then used to examine the dynamic evolution of the system during gate operation, as depicted in Fig. 4. The solid green line represents the applied driving pulse shape.

Regardless of whether the initial state is $|100\rangle$ or $|010\rangle$, a maximally entangled state is achieved at the end of evolution. Conversely, the other two computational states $|000\rangle$ and $|110\rangle$ ultimately revert to their initial states, as shown in Fig. 4(b). This observation further confirms the successful realization of the $\sqrt{i\text{SWAP}}$ gate operation. During the gate operation, the oscillation of quantum state $|000\rangle$ is almost negligible and states $|100\rangle$ or $|010\rangle$ exhibit slight oscillations. However, during the gate operation, the population of state $|110\rangle$ may transition to other states, but it ultimately returns to its initial state.

To assess the accuracy of a quantum gate operation, calculating its fidelity is essential. Next, we delve into the fidelity of quantum gates. As discussed in Appendix B, up to single-qubit phases, we numerically calculate the gate fidelity using Eq. (B1), and the fidelity of the $\sqrt{i\text{SWAP}}$ gate reaches 94.60\%. The slightly lower fidelity primarily stems from the accumulation of phase on state $|110\rangle$, despite compensating for the single-qubit gate phase. Further improvement in fidelity can be achieved by considering the cumulative phase of quantum state $|110\rangle$. Incorporating an arbitrary phase $\phi$ for state $|110\rangle$, the gate assumes a $\sqrt{i\text{SWAP}}$-like behavior. The conditional phase $\phi$ mainly arises from the resident ZZ coupling and microwave drives performed on the system. Considering the single-qubit phases and a conditional phase $\phi \approx 1.044$ rad, the fidelity of gate operation exceeds 99.92\%.


\begin{figure}
\begin{center}
\includegraphics[width=4.40cm, height=4.25cm]{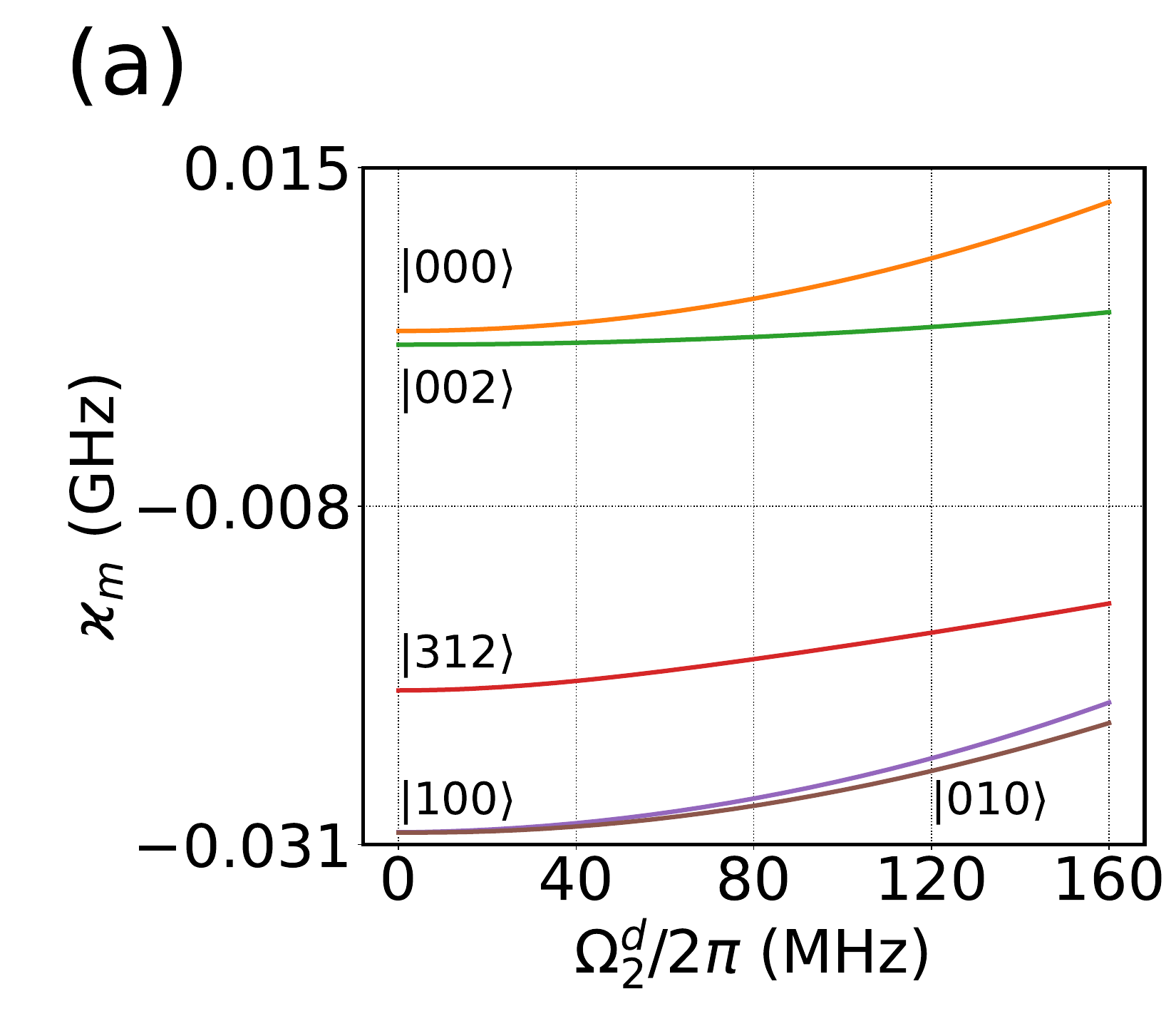}\includegraphics[width=4.10cm, height=4.15cm]{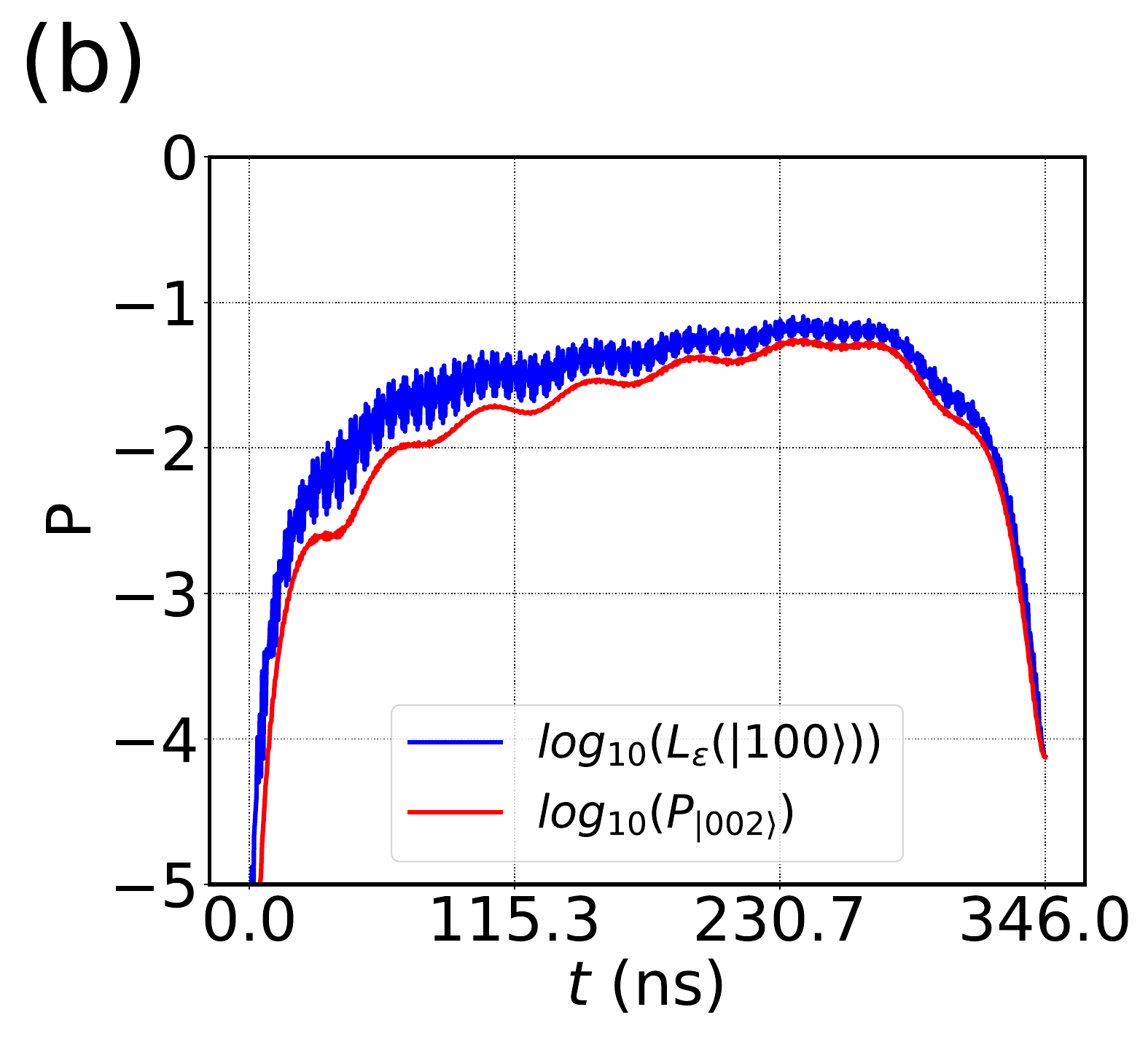}
\includegraphics[width=4.40cm, height=4.15cm]{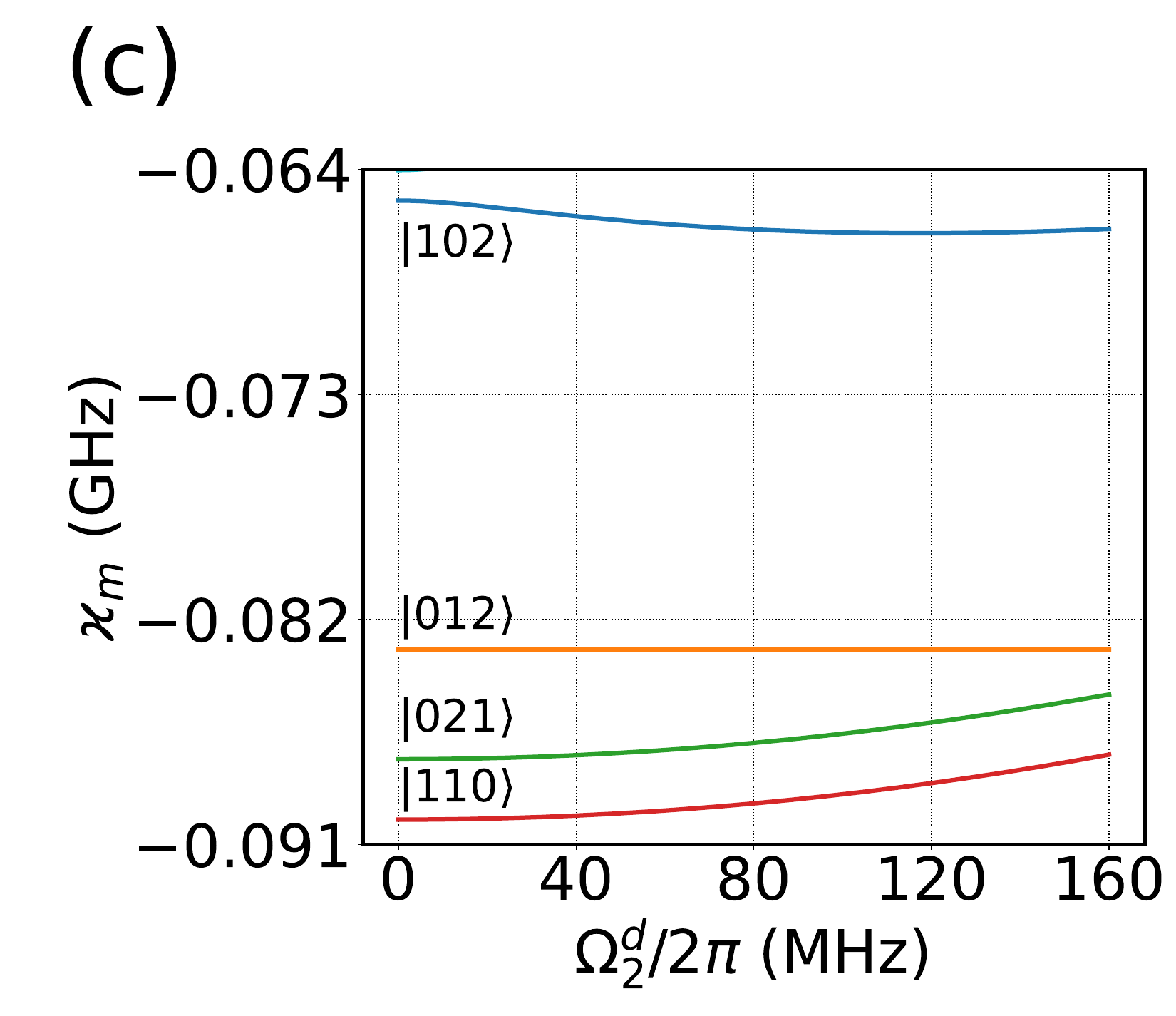}\includegraphics[width=4.10cm, height=4.10cm]{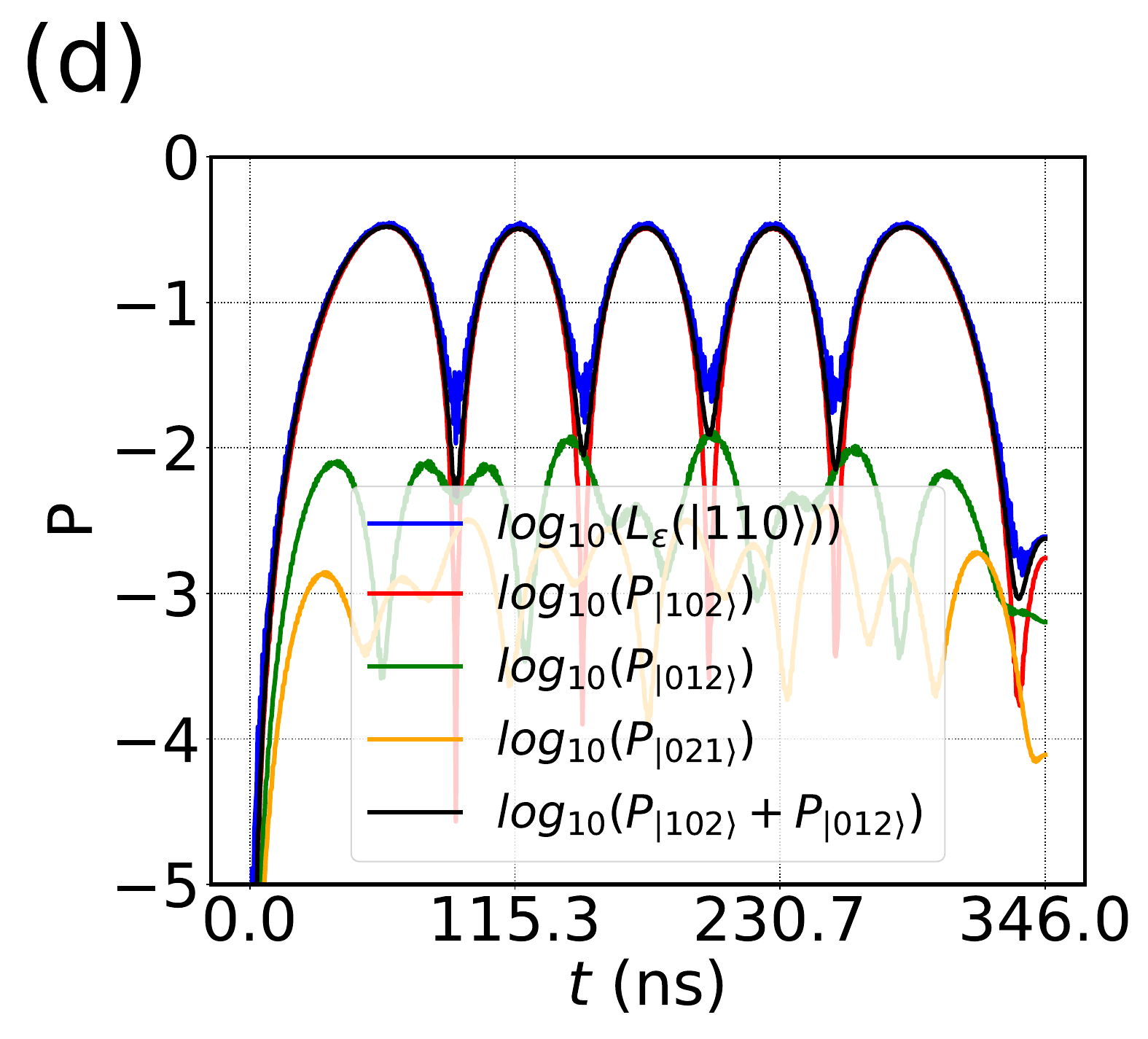}
\end{center}
\caption{{Numerical simulations for the quasienergies and leakage errors in the context of ABA architecture. Panel (a) depicts the quasienergies of state $|100\rangle$ (or $|010\rangle$) alongside neighboring states, including $|000\rangle$, $|002\rangle$, $|312\rangle$, $|100\rangle$, and $|010\rangle$, listed from top to bottom. In panel (b), the evolution of leakage errors (blue line) originating from state $|100\rangle$ is illustrated, alongside the population dynamics of the predominant leakage state $|002\rangle$ (red one) throughout the gate operation. Notably, The changes in leak errors are very similar to the changes in populations of state $|002\rangle$. Panel (c) presents the quasienergies for state $|110\rangle$ and its neighboring states, namely $|102\rangle$, $|012\rangle$, $|021\rangle$, and $|110\rangle$, listed in descending order. Among these, one state serves as the main leakage state. Panel (d) showcases the change in leakage errors (blue line) and the population dynamics of states $|102\rangle$ (in red), $|012\rangle$ (in green), and the combined population of $|102\rangle$ and $|012\rangle$ (in black) originating from state $|110\rangle$. Remarkably, the similar trends observed among these entities indicate that the primary leakage of state $|110\rangle$ is $|102\rangle$}.
\label{FigFive}}
\end{figure}

Additionally, leakage errors defined in Eq. (B2) significantly impact the fidelity of quantum gates, arising from unintended interactions between computational and non-computational bases. As depicted in Fig. 4 (b), the near-suppression of the $|000\rangle$ state's evolution results in a negligible leakage error after the gate operation. Our focus is on leakages from states $|100\rangle$ ($|010\rangle$) and $|110\rangle$. Utilizing Floquet theory (See Appendix B for details), we conduct numerical calculations of the quasienergies for states $|100\rangle$ (or $|010\rangle$) and $|110\rangle$, along with neighboring states that could engage in off-resonant interactions, as depicted in Fig. 5(a) and Fig. 5(c). In Fig. 5(a), it's evident that the leakage from states $|100\rangle$ and $|010\rangle$ corresponds to the same state $|002\rangle$, given the impulse drive between states $|100\rangle$ and $|002\rangle$. Conversely, state $|312\rangle$ engages in higher-order interactions with $|100\rangle$ (or $|010\rangle)$, and the probability of leakage to it is minimal. Further confirmation of the leakage state from $|100\rangle$ is provided by numerically calculating the leakage error $L_{\varepsilon}(|100\rangle)$ (blue) and the populations of leakage state $|002\rangle$ (red line) with the system initialized in $|100\rangle$, as shown in Fig. 5(b). Remarkably, their close alignment reinforces that the primary leakage state is indeed $|002\rangle$. Additionally, the similarity between the negligible leakage error and populations of state $|002\rangle$ after the gate operation further validates this finding. A similar analysis in Fig. 5(a) reveals that the primary leakage state from $|010\rangle$ is also $|002\rangle$.

In Fig. 5(c), there exist several states ($|102\rangle$, $|012\rangle$, $|021\rangle$) nearby the state $|110\rangle$ which could contribute the leakage. To identify the main leakage state from $|110\rangle$, we consider both energy differences and effective coupling strengths, revealing that the leakage state is $|102\rangle$. To validate this, we numerically simulate the change in leakage errors (blue) and populations of state $|102\rangle$ (red) during the gate operation process with the initial state $|110\rangle$, as depicted in Fig. 5(d). The synchronous fluctuations between the errors leaking from $|110\rangle$ and the change in $|102\rangle$ confirm that the maximal leakage state is indeed $|102\rangle$. Moreover, the calculated leakage error $L_{\varepsilon}(|110\rangle) \approx 2.5 \times 10^{-3}$ and populations $P(|102\rangle) \approx 1.8 \times 10^{-3}$ of the leakage state $|102\rangle$ after gate operation are consistent with Fig. 5(d), underscoring residual leakages in other quantum states after the gate operation. This is evidenced by plotting the populations of states $|012\rangle$ (green) and $P_{|102\rangle}$ + $P_{|012\rangle}$ (black) during the gate operation, with final populations approximately $0.6 \times 10^{-3}$ and $2.4 \times 10^{-3}$, respectively. Notably, the leakage error is approximately equal to the populations of states $P_{|102\rangle}$ + $P_{|012\rangle}$ after the gate operation, as illustrated in Fig. 5(d). Furthermore, accounting for the negligible leakage to state $|021\rangle$, it's evident that the leakage error $L_{\varepsilon}(|110\rangle)$ is indeed equal to the sum of populations (approximately $2.5 \times 10^{-3}$) for all three leakage states $|102\rangle$, $|012\rangle$, and $|021\rangle$ after the gate operation.


\subsection{ABC Architecture: Analysis and Implementation}

\begin{table}[h]
\centering
\caption{System parameters for two-qubit $\sqrt{i\text{SWAP}}$ gate in the ABC-type architecture.}
\label{my-label}
\begin{tabular}{@{}lccccc@{}}
\toprule
              & Bare frequency (GHz) & Anharmonicity (MHz)   & Coupling (MHz)  \\
\hline
$Q_1$ &  $\omega_1/2\pi$ = 5.0        &          $\alpha_1/2\pi = -300$    & \multicolumn{1}{c}{\multirow{2}{*}{$g_{1c}/2\pi$ = 110}}    \\
$Q_c$     & $\omega_c/2\pi$ =  6.2        &             $\alpha_c/2\pi = -400$      &   \multicolumn{1}{c}{\multirow{2}{*}{$g_{2c}/2\pi$ = 120}}  \\
$Q_2$ &  $\omega_2/2\pi$ = 7.5        &          $\alpha_2/2\pi = -300$    &                               \\  
\toprule
\end{tabular}
\end{table}

Next, we delve into the case of the ABC-type architecture with system parameters outlined in Table II. Considering the system parameters and the selected detuning $\delta/2\pi = -40$ MHz, the drive frequency is estimated as $\omega^d_1/2\pi \approx 6.979$ GHz. However, due to interactions between qubits and the coupler, the approximate value of the driving frequency $\omega^d_2$ will deviate from Eq. (6). As the $\sqrt{i\text{SWAP}}$ gate aims to create maximum entanglement between $|100\rangle$ and $|010\rangle$, we conduct numerical analysis on the populations of states $|100\rangle$ and $|010\rangle$, considering the pulse drive frequency $\omega^d_2$ and evolution time $t$, with an initial state of $|100\rangle$. By examining the population changes of these states, we accurately determine the drive frequency $\omega^d_2/2\pi \approx 4.458$ GHz.

\begin{figure}
\begin{center}
\includegraphics[width=4.60cm, height=4.05cm]{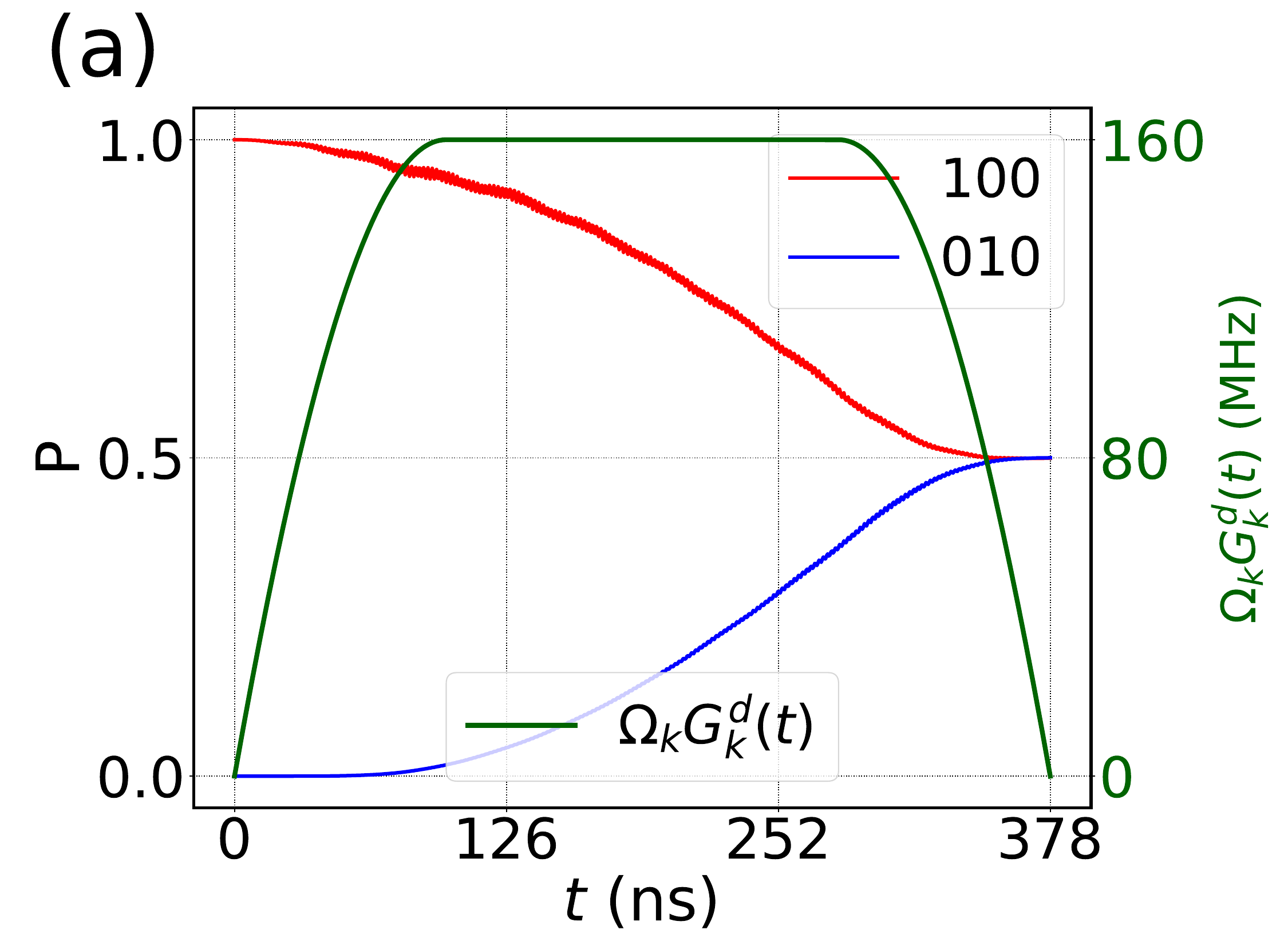}\includegraphics[width=4.0cm, height=4.05cm]{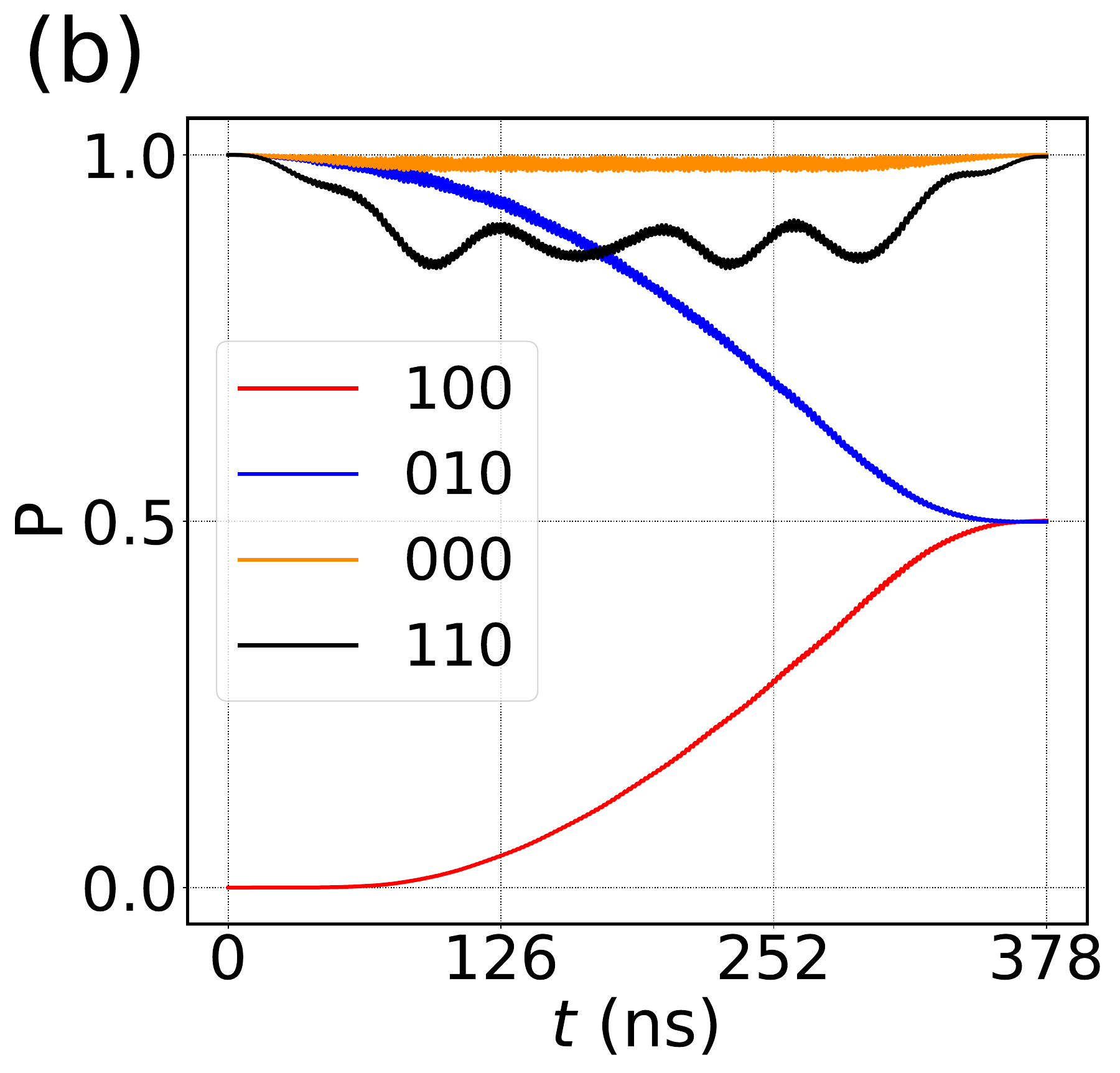}
\end{center}
\caption{The system dynamics during the gate operation for the ABC architecture. (a) The evolution of populations for states $|100\rangle$ (red) and $|010\rangle$ (blue) is depicted with the initial state $|100\rangle$. The green curve represents the time-dependent pulse shape with parameters $\gamma = 196$ ns and $t_p = 378$ ns. (b) Populations of states $|100\rangle$ ($|010\rangle$), $|000\rangle$ (orange), and $|110\rangle$ (black) are shown corresponding to different initial states $|010\rangle$, $|000\rangle$, and $|110\rangle$, respectively. The pulse parameters are set to $\Omega_{1(2)}(0)/2\pi \approx 160$ MHz, $\omega^d_1/2\pi \approx 6.979$ GHz, $\omega^d_2/2\pi \approx 4.458$ GHz, with a fixed relation $t_r = 0.5\gamma$, and the system parameter values are presented in Table II.
\label{FigSix}}
\end{figure}

During the operation of quantum gates, leakage errors will occur. However, this leakage error can be suppressed when using a flat waveform. Considering the time-dependent pulse shape $G^{d}_{k}(t)$ in Eq. (8), can effectively suppress leakages induced by off-resonance interaction associated with non-computational basis states. For simplicity, we maintain the relationship $ t_r = 0.5\gamma$ and conduct numerical calculations for the populations of computational basis states as a function of parameter $\gamma$ and time $t$. Based on the populations of states $|000\rangle$, $|100\rangle$, $|010\rangle$, and $|110\rangle$, we select specific parameter values: $\gamma = 196$ ns and $t=378$ ns. Considering the shape of the driving pulse and the chosen parameters, we simulate the system dynamics during the gate operation, as depicted in Fig. 6. Regardless of whether the initial state is $|100\rangle$ or $|010\rangle$, a maximum entangled state is eventually achieved. Additionally, the quantum states $|000\rangle$ and $|110\rangle$ revert to their initial states, indicating the successful implementation of a $\sqrt{i\text{SWAP}}$ gate operation.



Next, we calculate the fidelity of the gate according to the definition in Eq. (B1). Following the inclusion of single-qubit phase compensations, the fidelity of the $\sqrt{i\text{SWAP}}$ gate surpasses 99.63\%. This represents a significant improvement compared to the ABA-type architecture, primarily due to the substantial suppression of static ZZ interaction in the ABC-type architecture (approximately $-36$ kHz). Considering the phase $\phi \approx 0.246$ rad of state $|110\rangle$, the fidelity of the $\sqrt{i\text{SWAP}}$-like gate implemented at this stage exceeds 99.93\%. Examining the dynamical evolution of the system in Fig. 6, it is evident that states $|100\rangle$ and $|010\rangle$ exhibit minimal quantum oscillations during the gate operation. The state $|000\rangle$ remains nearly unchanged, while state $|110\rangle$ experiences significant quantum oscillations.

\begin{figure}
\begin{center}
\includegraphics[width=4.20cm, height=4.20cm]{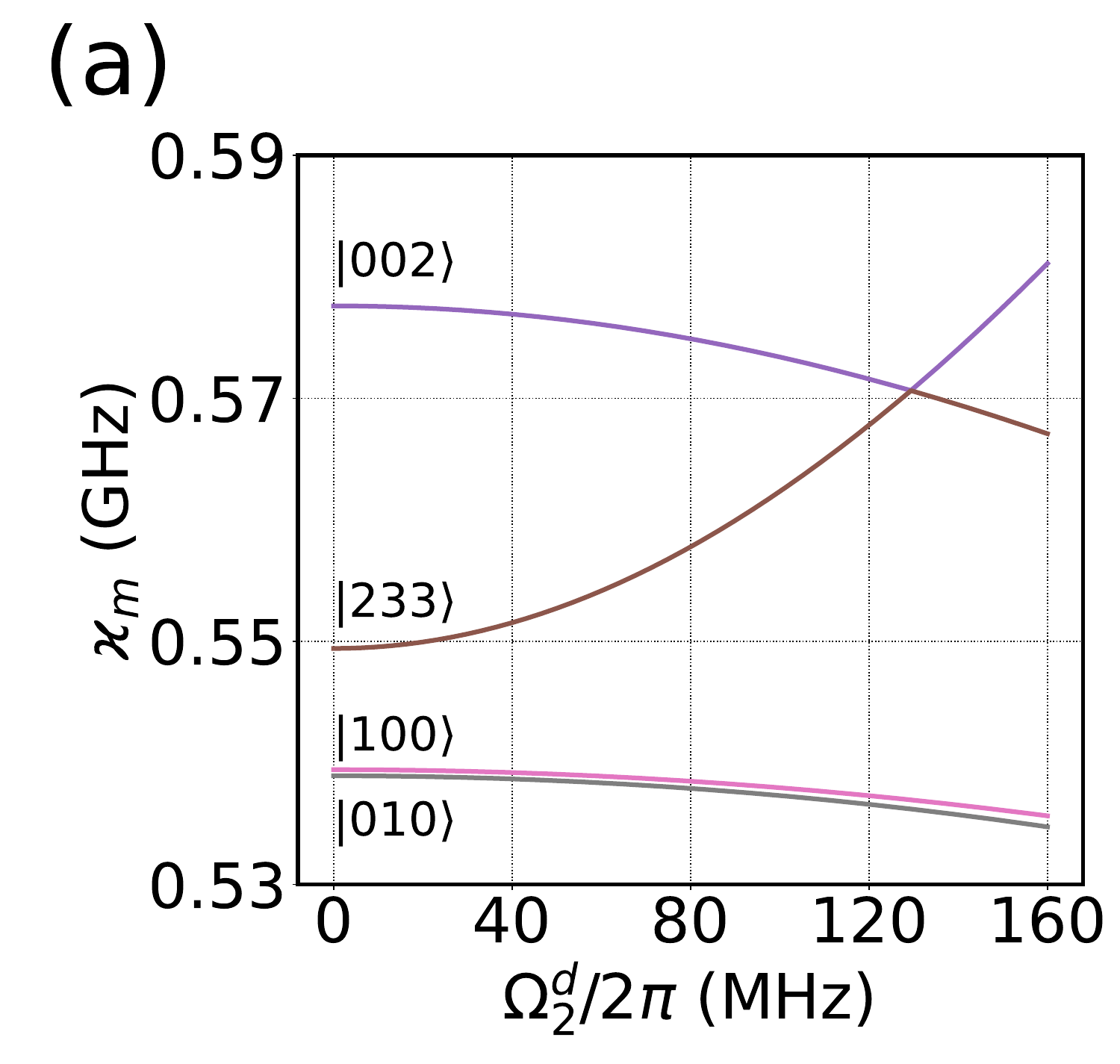}\includegraphics[width=4.20cm, height=4.20cm]{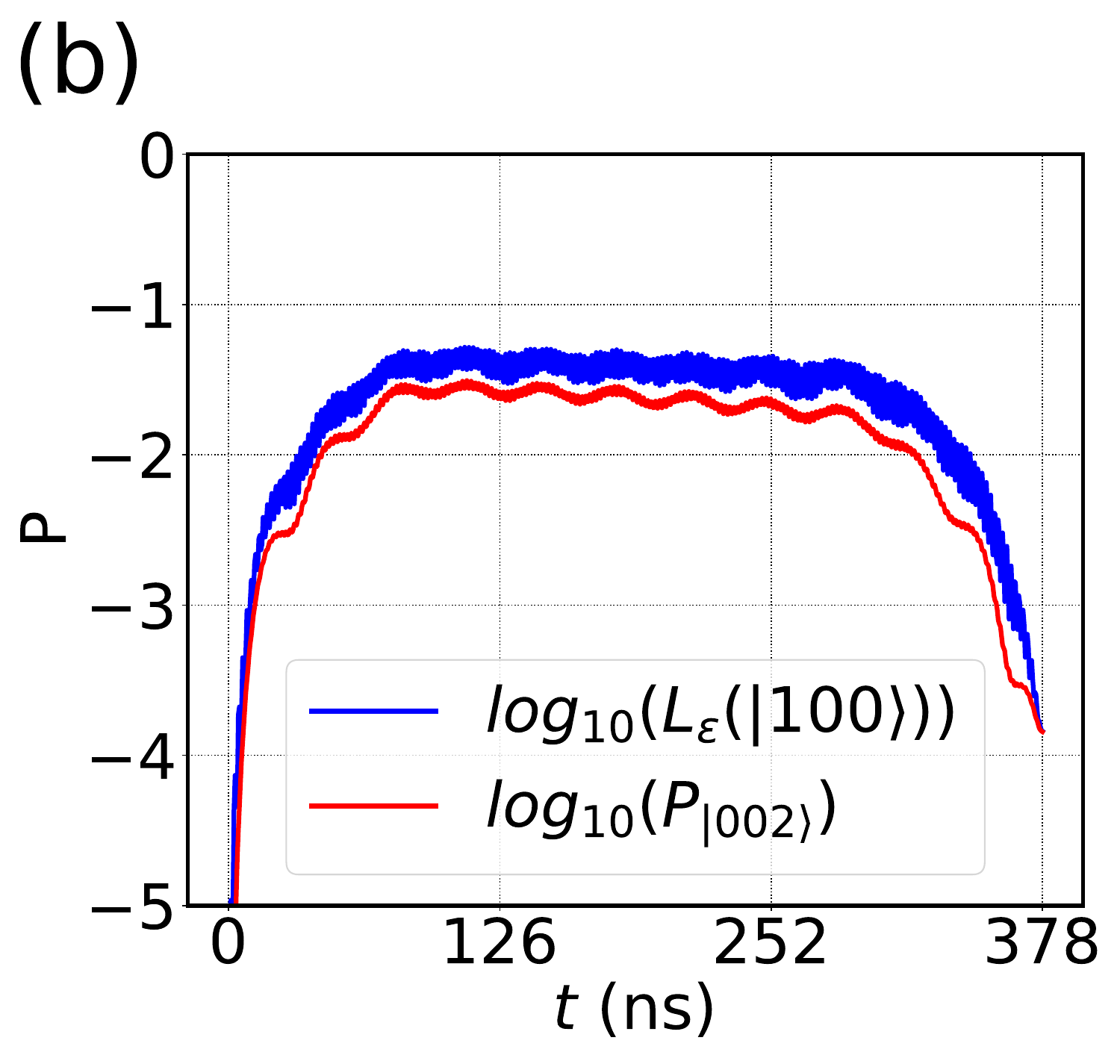}
\includegraphics[width=4.20cm, height=4.20cm]{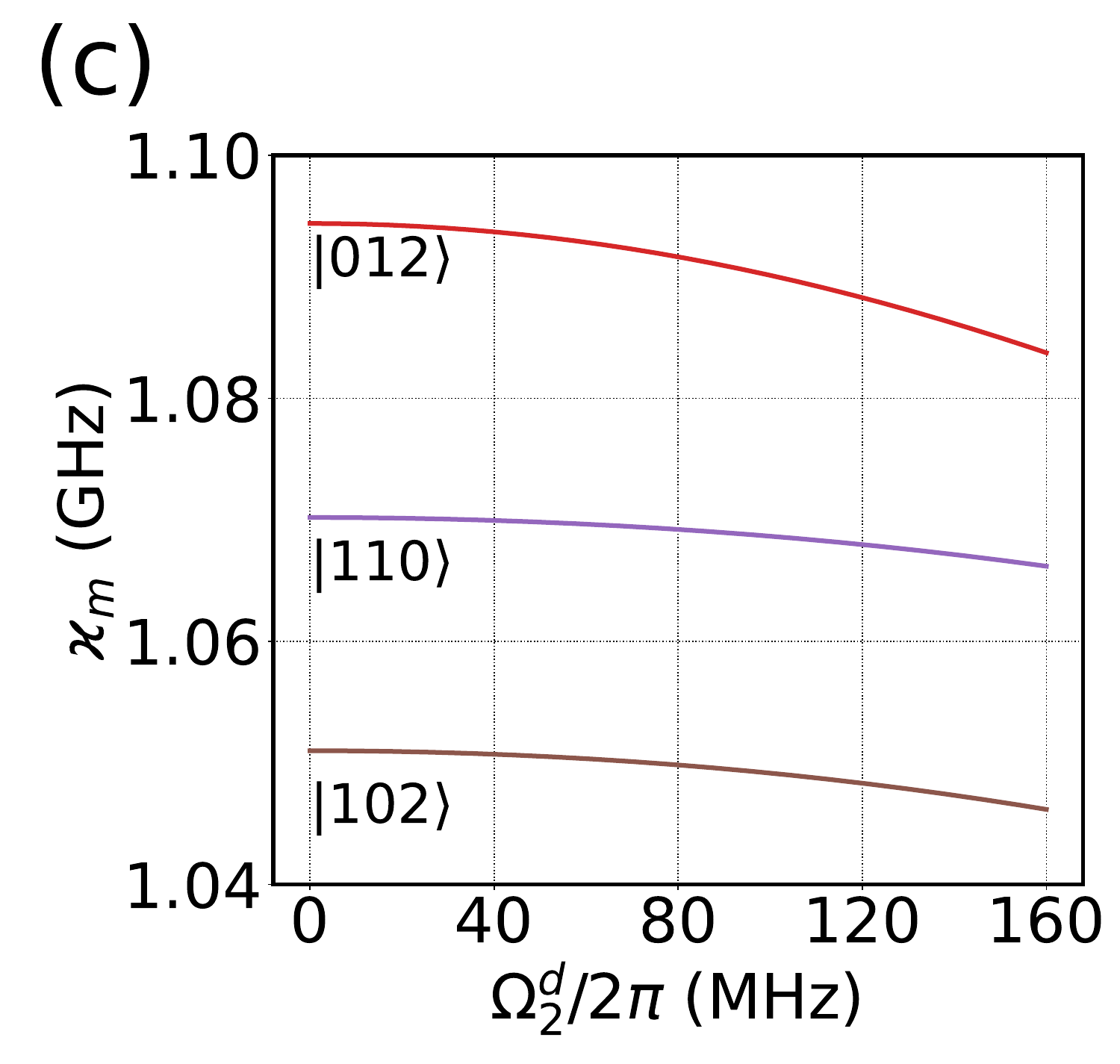}\includegraphics[width=4.20cm, height=4.20cm]{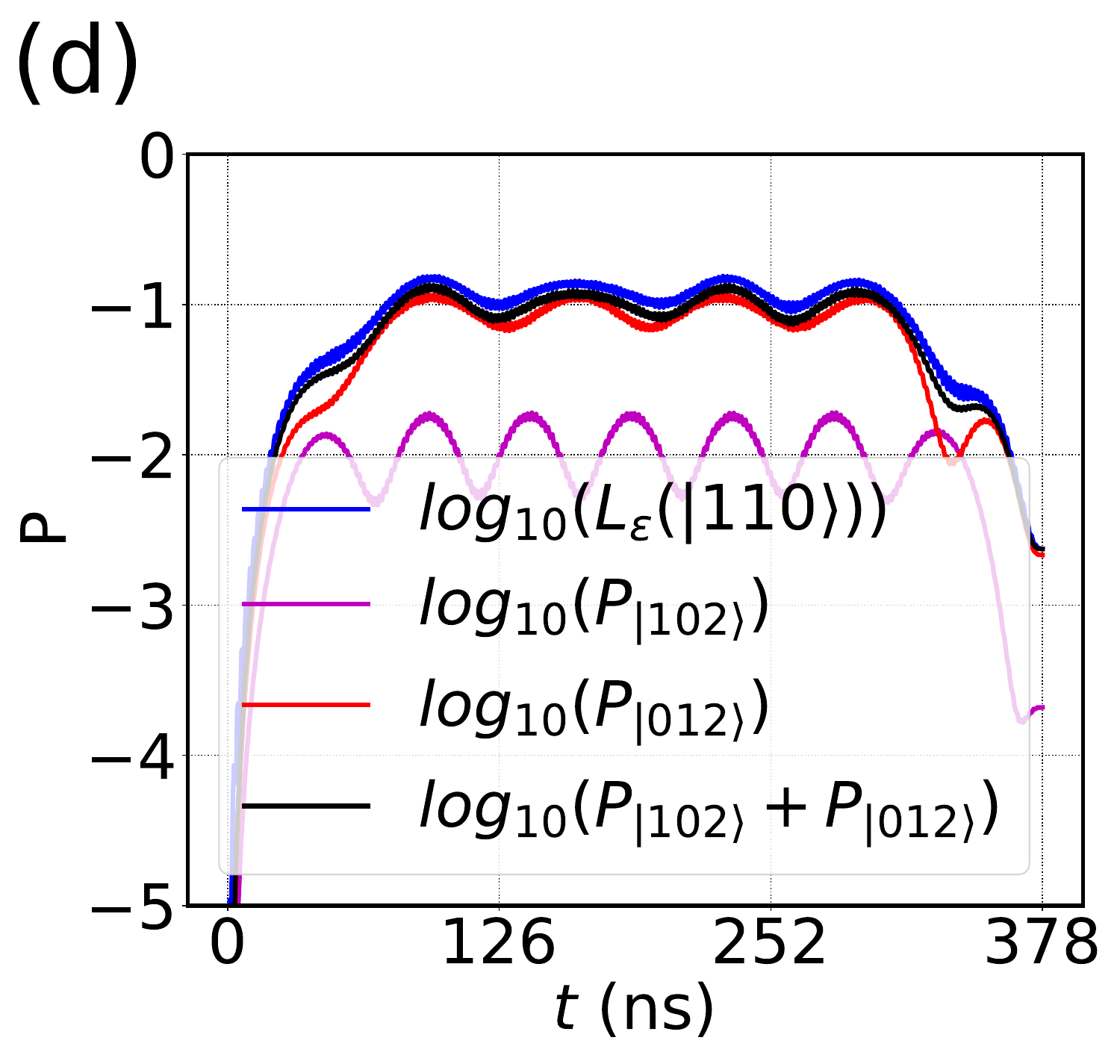}
\end{center}
\caption{{The quasienergies and leakage errors in the ABC-type architecture. (a) Quasienergies surrounding state $|100\rangle$, including $|002\rangle$, $|233\rangle$, $|100\rangle$, and $|010\rangle$, are depicted from top to bottom. The most significant leakage state likely originates from one of them. (b) Presents the calculation results of leakage errors (blue) from $|100\rangle$ alongside populations (red) of state $|002\rangle$. The consistent trends between them confirm that $|002\rangle$ is the primary leakage from $|100\rangle$. (c) Considering state $|110\rangle$, quasienergies proximal to it, namely $|012\rangle$, $|110\rangle$, and $|102\rangle$. (d) Numerical assessments of the leakage error (blue) from $|110\rangle$ are juxtaposed with populations of state $|102\rangle$ (magenta), $|012\rangle$ (red), and their sum $P_{|102\rangle}$ + $P_{|012\rangle}$ (black). The similar trends between the blue and red curves indicate that the primary leakage state is $|012\rangle$}.
\label{FigSeven}}
\end{figure}

Following the discussions in Appendix B, employing the Floquet theory enables us to computationally determine the quasienergies surrounding states $|100\rangle$ (or $|010\rangle$) and $|110\rangle$, as depicted in Figs. 7(a,c). We aim to identify the primary leakage state among them. In Fig. 7(a), both $|002\rangle$ and $|233\rangle$ could potentially be the primary leakage state. However, the interaction with state $|233\rangle$ constitutes a higher-order process, minimizing the likelihood of leakage into it. Therefore, it becomes evident that the most probable leakage of state $|100\rangle$ is $|002\rangle$. To confirm this, we further analyze the leakage state $|002\rangle$ (rather than $|233\rangle$) by numerically evaluating the variations in leakage errors and populations for state $|002\rangle$, as displayed in Fig. 7(b). The results indicate that the leakage errors from state $|100\rangle$ exhibit almost similar changes in $|002\rangle$ throughout the dynamics. Notably, the leakage error $L_{\varepsilon}(|100\rangle)$ is approximately $0.1 \times 10^{-3}$, nearly identical to the populations of state $|002\rangle$ after the gate operation, confirming that the primary leakage from state $|100\rangle$ is indeed $|002\rangle$. Similarly, according to Fig. 7(a), the main leakage from $|010\rangle$ is also state $|002\rangle$, with ignorable leakage errors almost equivalent to the populations of state $|002\rangle$ after the gate operation.

As depicted in Fig. 7(c), the leakage from $|110\rangle$ potentially involves $|012\rangle$ or $|102\rangle$. To ascertain the precise state, we plot the leakage error $L_{\varepsilon}(|110\rangle)$, along with populations of states $|102\rangle$, $|012\rangle$, and their sum $P_{|102\rangle}$ + $P_{|012\rangle}$ in Fig. 7(d). The similarity between the leakage error $L_{\varepsilon}(|110\rangle)$ and populations of state $|012\rangle$ further corroborates that the primary leakage from state $|110\rangle$ predominantly involves $|012\rangle$. After the gate operation, we observe that the leakage error $L_{\varepsilon}(|110\rangle)$ is approximately $2.4 \times 10^{-3}$, closely aligned with the populations of state $|012\rangle$ and nearly equivalent to the sum of populations for states $|102\rangle$ and $|012\rangle$, as shown in Fig 7(d).

\begin{figure}
\begin{center}
\includegraphics[width=4.20cm, height=4.20cm]{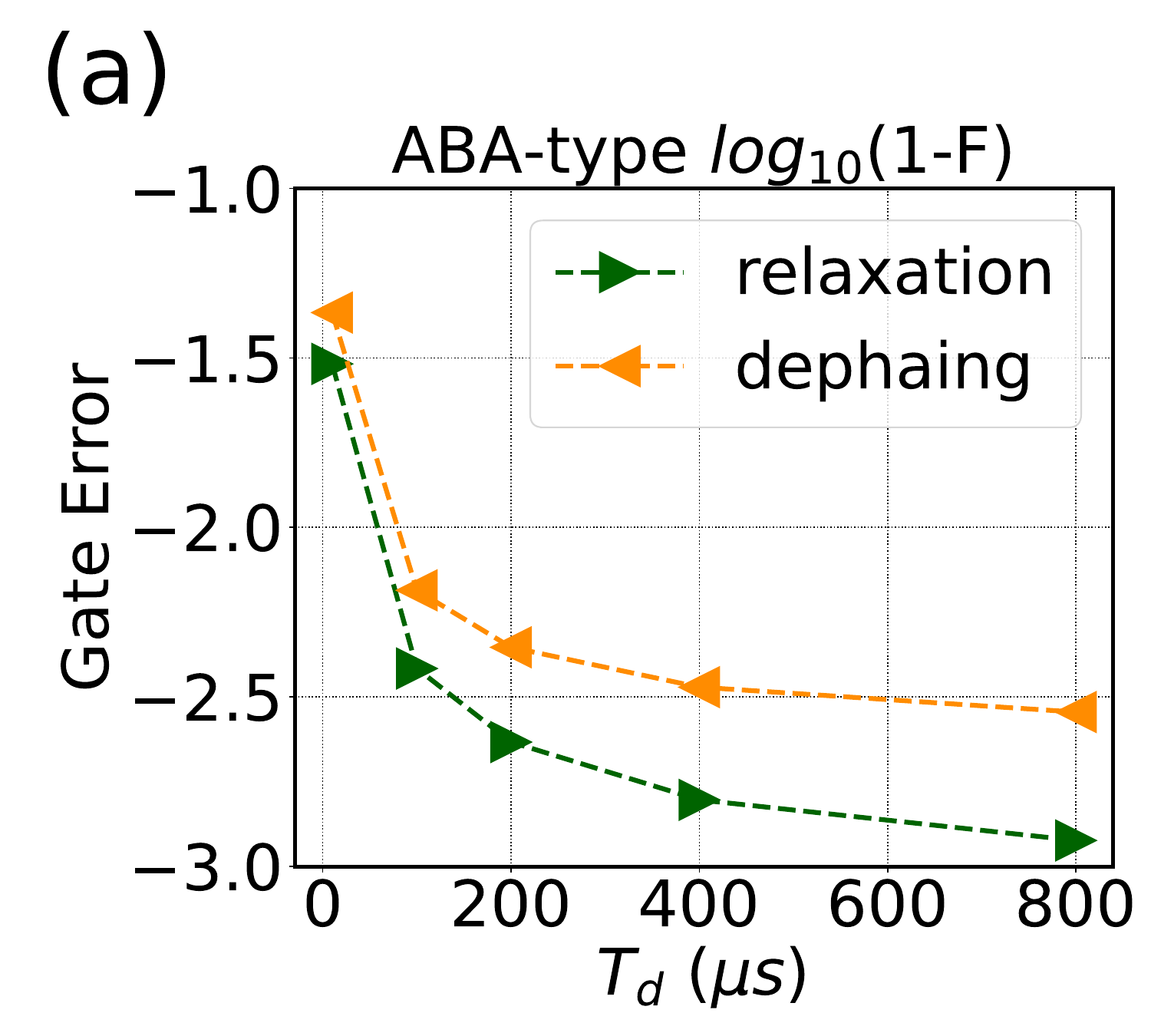}\includegraphics[width=4.20cm, height=4.20cm]{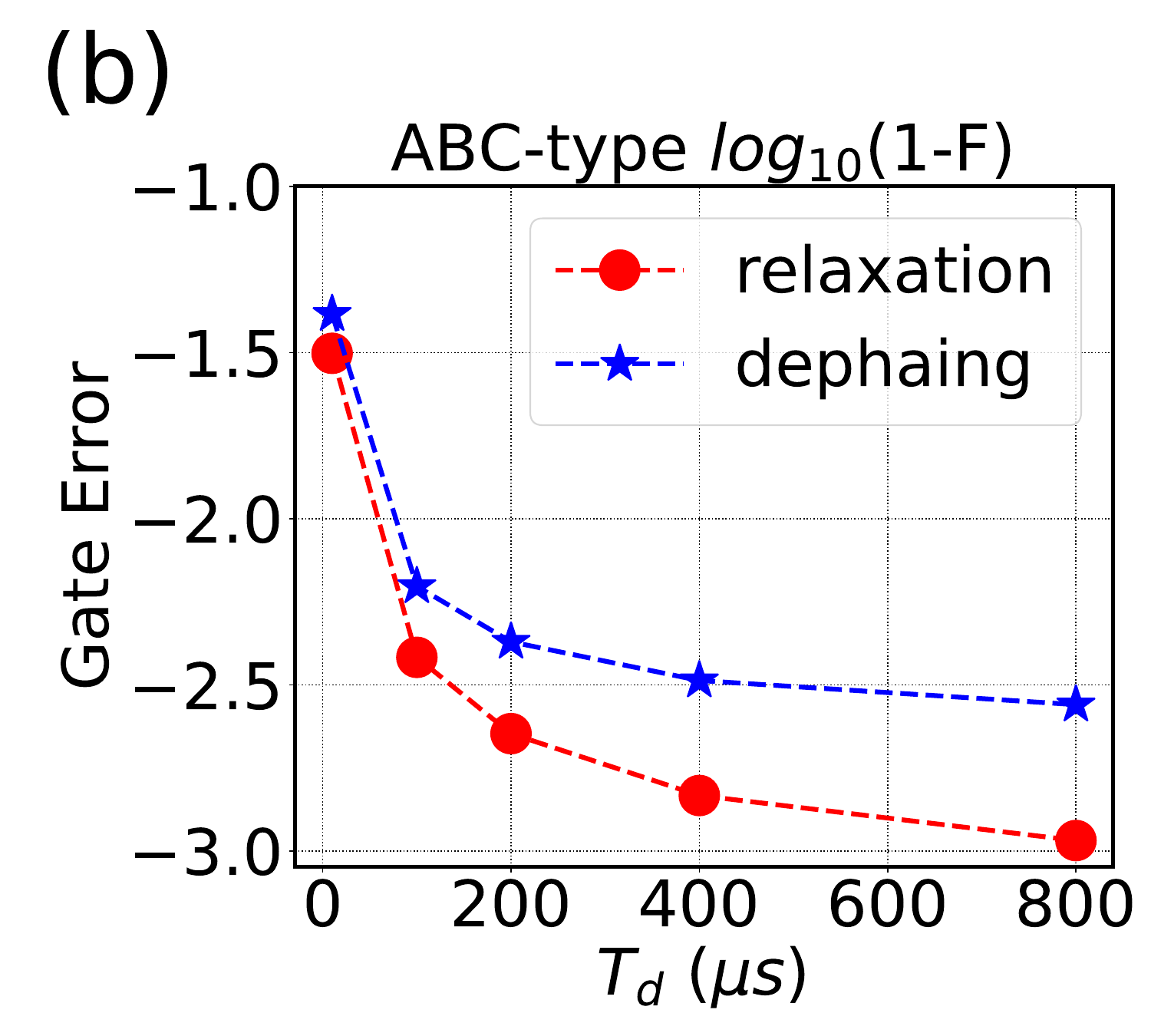}
\end{center}
\caption{{The gate error under the decoherence processes. In (a), the gate error is plotted against qubit decoherence times for the ABA-type architecture. Due to the limitation of qubit relaxation time $T_{1,2,c}$, the gate fidelity (triangle right) increases as the coherence time extends. The variation in gate error (triangle left) is depicted for different qubit dephasing times $T_{\phi}$ while maintaining a fixed qubit relaxation time $T_{1,2,c}$ = 200 $\mu$s. In (b), the gate error is illustrated as a function of decoherence processes in ABC-type architecture. As qubit relaxation time decreases, the gate error (circle) increases. With a fixed qubit relaxation time $T_{1,2,c}$ = 200 $\mu$s, the gate error (star) gradually decreases with an increase in qubit phase time. It is assumed that the transmon qubits and coupler possess identical relaxation and dephasing times.}
\label{FigEight}}
\end{figure}

\section{Discussions}

\subsection{The effect of decoherence}

The gate performance described above does not account for the system decoherence process. Next, we examine the impact of relaxation on gate performance. Two primary sources of gate errors arise from the relaxation and dephasing processes of the transmon qubits. To address qubit decoherence, we employ the following master equation \cite{PengZhaoAPL2022, ChenYinqiPRAp2022} to simulate the system dynamics
\begin{equation}
\begin{aligned}
\dot{\rho}(t) = &-i[H_{s} + H_{d},\rho] + \sum_{j}\mathcal{D}[\mathcal{C}_j]\rho,
\end{aligned}
\end{equation} 
with $\mathcal{D}[\mathcal{C}]\rho = \mathcal{C}\rho \mathcal{C}^{\dagger} - (\mathcal{C}^{\dagger}\mathcal{C}\rho + \rho\mathcal{C}^{\dagger}\mathcal{C})/2$ and $\mathcal{C}_{j}$ $ = \{\sqrt{1/T_{j}}q_{j}, \sqrt{2/T_{p_{j}}}q^{\dagger}_{j} q_{j} \}$. Here, $T_{j}$ and $T_{p_{j}}$ represent the relaxation and dephasing times of qubit-$j$.

For simplicity, we assume uniform relaxation rates across all transmon qubits (coupler) and numerically evaluate the fidelity of the $\sqrt{i\text{SWAP}}$-like gate using Eq. (B1) by simulating the aforementioned master equation. The results for both ABA and ABC-type architectures are depicted in Fig. 8. Notably, gate error is influenced by decoherence. Shorter coherence times lead to higher error rates. Figure 8(a) illustrates the variation of gate error concerning relaxation (triangle right) and dephasing (triangle left) times for ABA-type qubits. With short qubit relaxation times $T_{1,2,c}$ = 10 $\mu$s, the gate error substantially elevates to 0.0304. Conversely, the error reduces to below 0.0012 when relaxation time exceeds 800 $\mu$s. In the dephasing scenario, the triangle left data represent fixed qubit relaxation times $T_{1,2,c}$ = 200 $\mu$s, showing a gate error of 0.0430 for a low qubit dephasing time 10 $\mu$s. However, gate fidelity improves to 0.9971 with dephasing times exceeding 800 $\mu$s.

In the ABC-type scenario, the results illustrate similar changes in gate error under qubit decoherence, as shown in Fig. 8(b). The gate error peaks at 0.0314 with qubit relaxation times of $T_{1,2,c}$ = 10 $\mu$s. Conversely, a high-fidelity $\sqrt{i\text{SWAP}}$-like gate with an error below 0.0015 can be achieved with superconducting qubits' relaxation times exceeding 800 $\mu$s. Considering the constraint of qubit relaxation time $T_{1,2,c}$ = 200 $\mu s$, the gate error is approximately 0.0412 for a qubit dephasing time of 10 $\mu$s. However, the gate error is suppressed below 0.0028 for a longer dephasing time of 800 $\mu$s. With advancements in modern technology, state-of-the-art experiments in superconducting quantum systems report decoherence times \cite{YuHaifeng2022, PlaceNC2021} that can far exceed the gate operation time.

\subsection{The ZZ coupling}

Notably, disregarding the additional phase introduced by the ZZ interaction, the fidelity of the $\sqrt{i\text{SWAP}}$-like gate achieved in both ABA and ABC-type architectures can exceed 99.92\%. Expanding on the preceding discussion, distinct qubit energy-level architectures yield varied ZZ interactions. During the realization of $\sqrt{i\text{SWAP}}$ gate, we observe that the static ZZ coupling is approximately $-273$ kHz in the ABA-type architecture. Despite incorporating single-qubit phase compensation operations, the fidelity of the implemented $\sqrt{i\text{SWAP}}$ gate remains suboptimal, resulting in a fidelity of 94.60\%. The main reason is that the static ZZ interaction is relatively large, and the fidelity of the $\sqrt{i\text{SWAP}}$ gate operation is impacted by the conditional phase of state $|110\rangle$. Conversely, in an ABC-type energy level structure, the ZZ coupling has been suppressed to approximately $-36$ kHz, leading to notably higher fidelity of the implemented $\sqrt{i\text{SWAP}}$ gate, reaching up to 99.63\%. A comparison between ABA and ABC setups suggests that designing qubit energy-level architectures can effectively mitigate ZZ interaction and significantly enhance the fidelity of the $\sqrt{i\text{SWAP}}$ gate. This underscores the considerable impact of the phase introduced by the static ZZ interaction on the gate's fidelity. Additionally, it is important to note that the phase is not solely caused by the static ZZ interaction but also by the dynamic ZZ interaction induced by external drivings \cite{FeiYanPRL129040502, NakamuraPRL2606012023}.

\section{Conclusion}

In this study, we introduce a microwave-driven approach to implement the $\sqrt{i\text{SWAP}}$ gate using fixed-frequency superconducting qubits. Two qubits are connected via a shared transmon coupler with a constant interaction strength, with microwave drives applied solely to the coupler. By carefully designing the driving pulse profile and selecting appropriate parameters for the coupler-driven pulses based on the system dynamics, we can enhance the fidelity of gate operations. Further improvements in gate fidelity and operational timing can be achieved by optimizing the quantum system's parameter values and the characteristics of the microwave pulses. Additionally, we assess the impact of qubit decoherence and determine that longer relaxation times can sufficiently mitigate gate errors for both ABA- and ABC-type qubit architectures. Superconducting quantum devices have sparked optimism for large-scale and intricate quantum computing, owing to their capability to perform quantum operations across various qubits within multi-qubit systems. We demonstrate that the gate scheme can be effectively extended to multi-qubit systems, as shown through its application in a three-qubit setup. We discuss the implementation of $\sqrt{i\text{SWAP}}$-like gate operations between $Q_1$ and $Q_{2(3)}$, achieving high fidelity.

Therefore, this proposed gate scheme has the potential for constructing high-fidelity multibody gate operations. Furthermore, it can be adapted to various types of superconducting qubits and cases with different-sign anharmonicity. We anticipate that the applications of this protocol will facilitate the construction and streamlining of complex superconducting quantum computing systems.

\section*{Acknowledgments}

P. X would like to thank Peng Zhao for many helpful discussions on this work. Additionally, P. X expresses gratitude to Yumei Song for her steadfast support throughout my academic journey and acknowledges the Niels Bohr Institute for their hospitality during the visit, during which the majority of this work was conducted. Financial support for this research was provided by the National Natural Science Foundation of China (Grant Nos. 12105146, 12175104), along with a fellowship from the China Scholarship Council. S. W. also acknowledges funding from the Innovation Program for Quantum Science and Technology (2021ZD0301701) and the National Key Research and development Program of China (No. 2023YFC2205802).

\appendix

\section{The evolution paths}

Here, based on the perturbation theory, we give the derivation of the effective interaction between states $|100\rangle$ and $|010\rangle$. Following fourth-order perturbation theory, the effective interaction strength for two states ($|f\rangle$ and $|i\rangle$) could be written as \cite{AFKockumPRA0638492017, ZhaoPengPRA952017, PengZhaoPRA960438332017}
\begin{equation}
\begin{aligned}
J = \sum_{s_1,s_2,s_3}\frac{H_{i,s_1}H_{s_1,s_2}H_{s_2,s_3}H_{s_3,f}}{(E_{i} - E_{s_1})(E_{i} - E_{s_2})(E_{i} - E_{s_3})}.
\end{aligned}
\end{equation}
where $s_i$ indicates the mediate states connected states $|f\rangle$ and $|i\rangle$, $E_{s_i}$ is the energy of the state, and $H_{i,si}$ = $\langle i|H|s_i\rangle$. The sum goes over all of the virtual transition steps. In the following discussions, we consider that the two driving pulses are traded as two quantum modes. Therefore,  the full system states are denoted as $|Q_1, Q_2, Q_c, D_1, D_2\rangle$, which correspond to the states of qubit-1, qubit-2, coupler, and the two drive modes applied to the coupler, respectively \cite{XuPengPRA1082023}. For the proposed gate scheme, eight evolutionary paths linking states $|100\rangle$ and $|010\rangle$ are identified,
\begin{equation}
\begin{aligned}
&|10011\rangle \leftrightarrow |00111\rangle \leftrightarrow |00201\rangle \leftrightarrow |01101\rangle  \leftrightarrow |01002\rangle,&\\
&|10011\rangle \leftrightarrow |00111\rangle \leftrightarrow |00201\rangle \leftrightarrow |00102\rangle  \leftrightarrow |01002\rangle,&\\ 
&|10011\rangle \leftrightarrow |10101\rangle \leftrightarrow |00201\rangle \leftrightarrow |01101\rangle  \leftrightarrow |01002\rangle,&\\
&|10011\rangle \leftrightarrow |10101\rangle \leftrightarrow |00201\rangle \leftrightarrow |00102\rangle  \leftrightarrow |01002\rangle,& \\
&|10011\rangle \leftrightarrow |10101\rangle \leftrightarrow |11001\rangle \leftrightarrow |01101\rangle  \leftrightarrow |01002\rangle,&\\
&|10011\rangle \leftrightarrow |10101\rangle \leftrightarrow |10002\rangle \leftrightarrow |00102\rangle  \leftrightarrow |01002\rangle,&\\ 
&|10011\rangle \leftrightarrow |00111\rangle \leftrightarrow |01011\rangle \leftrightarrow |01101\rangle  \leftrightarrow |01002\rangle,&\\
&|10011\rangle \leftrightarrow |00111\rangle \leftrightarrow |00012\rangle \leftrightarrow |00102\rangle  \leftrightarrow |01002\rangle.&
\end{aligned}
\end{equation}
 According to Eq. (A1) and Eq. (A2), we can derive the effective coupling strength $J_{12}$ given in Eq. (7) of the main text.


\section{Analysis of Fidelity and Leakage in Quantum Systems}

Here, we will investigate the gate fidelity and leakage errors during the gate operation. The discussion on the implementation of $\sqrt{i\text{SWAP}}$ gates has been previously addressed. Evaluating the precision of the gate operation requires calculating the gate fidelity. To assess the performance of the implemented $\sqrt{i\text{SWAP}}$ gate, we employ the state-average gate fidelity metric, as defined in \cite{PedersenPLA2007}
\begin{equation}
\begin{aligned}
F=\frac{[Tr(U^{\dag}_{\text{real}}U_{\text{real}})+|Tr(U^{\dag}U_{\text{real}})|^2]}{20}.
\end{aligned}
\end{equation} 
The operator $U_{\text{real}}$ represents the actual evolution projected onto the computational subspace, evaluated using the full system Hamiltonian $H_s$ + $H_d$ (without considering decoherence effects). On the other hand, $U$ denotes the ideal $\sqrt{i\text{SWAP}}$ gate. It is important to note that during the system dynamics, different quantum states accumulate varying phases. To ensure the accuracy of multi-qubit gate operations, single-qubit phase errors can be compensated for through single-qubit gate operations, thereby enhancing the accuracy of the $\sqrt{i\text{SWAP}}$ gate. By leveraging the system dynamics, we can determine single-qubit phases through numerical optimization methods. For the single-qubit phases, corrective single-qubit phase operations \cite{GhoshPRA2013, ZahedinejadPRAp2016, BarnesPRB2017} can be implemented just before and after the $\sqrt{i\text{SWAP}}$ gate to rectify them.


In addition, leakage errors are widely recognized as a primary source of error that impacts gate operations significantly. These errors occur when states transition from the designated computational subspace to the noncomputational subspace during system dynamics. The computational subspace, denoted as $V_1$, encompasses the system's energy levels {$|00\rangle$, $|10\rangle$, $|01\rangle$, $|11\rangle$}, where ideal dynamics occur, with a dimensionality characterized by $d_1=4$. In the current gate scheme, all qubits are truncated to four levels, resulting in a 64-dimensional complete state space for the system. Leakage errors can be defined as follows \cite{JayMGambettaPRA970323062018},
\begin{equation}
\begin{aligned}
L_{\varepsilon} = 1- Tr[\frac{M_1V_1}{d_1}].
\end{aligned}
\end{equation}  
Leakage errors quantify the extent of transition from the computational to the noncomputational basis space, yet they do not precisely identify the leaked quantum states. To discern the leaked states more accurately, we can numerically compute the eigenstates of $H_F(t)$, derived with rotating wave approximation (RWA) and consideration of a rotating coordinate system with the driving pulse frequency $\omega^d_1$. These eigenstates, commonly known as Floquet states \cite{JHShirleyB9791965, YBZelDovichJETP10061967, VIRitusJETP1967, HSambePRA722031973}, provide insights into the system's behavior under the periodic nature of the driving pulse. Notably, we express the Hamiltonian $H_F(t)$ as $H_F(t + 2\pi/\nu)$, where $\nu = \omega^d_{2} - \omega^d_{1}$ represents the disparity between the two driving pulses. Subsequently, we employ the following procedure to determine the eigenstates,
\begin{equation}
\begin{aligned}
\text{[}H_F(t) - i\partial_t \text{]}|\Psi_m(t)\rangle = \varkappa_m|\Psi_m(t)\rangle.
\end{aligned}
\end{equation}
The solutions derived from Eq. (B3) provide quasienergies denoted as $\varkappa_m$, accompanied by corresponding eigenvectors representing the Floquet modes, where $m$ ranges from 1 to 64. It is crucial to highlight that the solutions within Eq. (B3) are constrained within integer multiples of the drive frequency $\nu$. In essence, if {${\varkappa_m,|\Psi_m(t)\rangle}$} constitutes a valid solution, then {${\varkappa_{mk} \equiv \varkappa_m + k\nu, |\Psi_{mk}(t)\rangle = e^{-i\nu t}|\Psi_m(t)\rangle}$} also qualifies as a solution, directly attributable to the periodic nature exhibited by the Floquet modes.

\begin{figure}
\vspace{0.0 em}  
\begin{center}
\includegraphics[width=4.60cm, height=4.05cm]{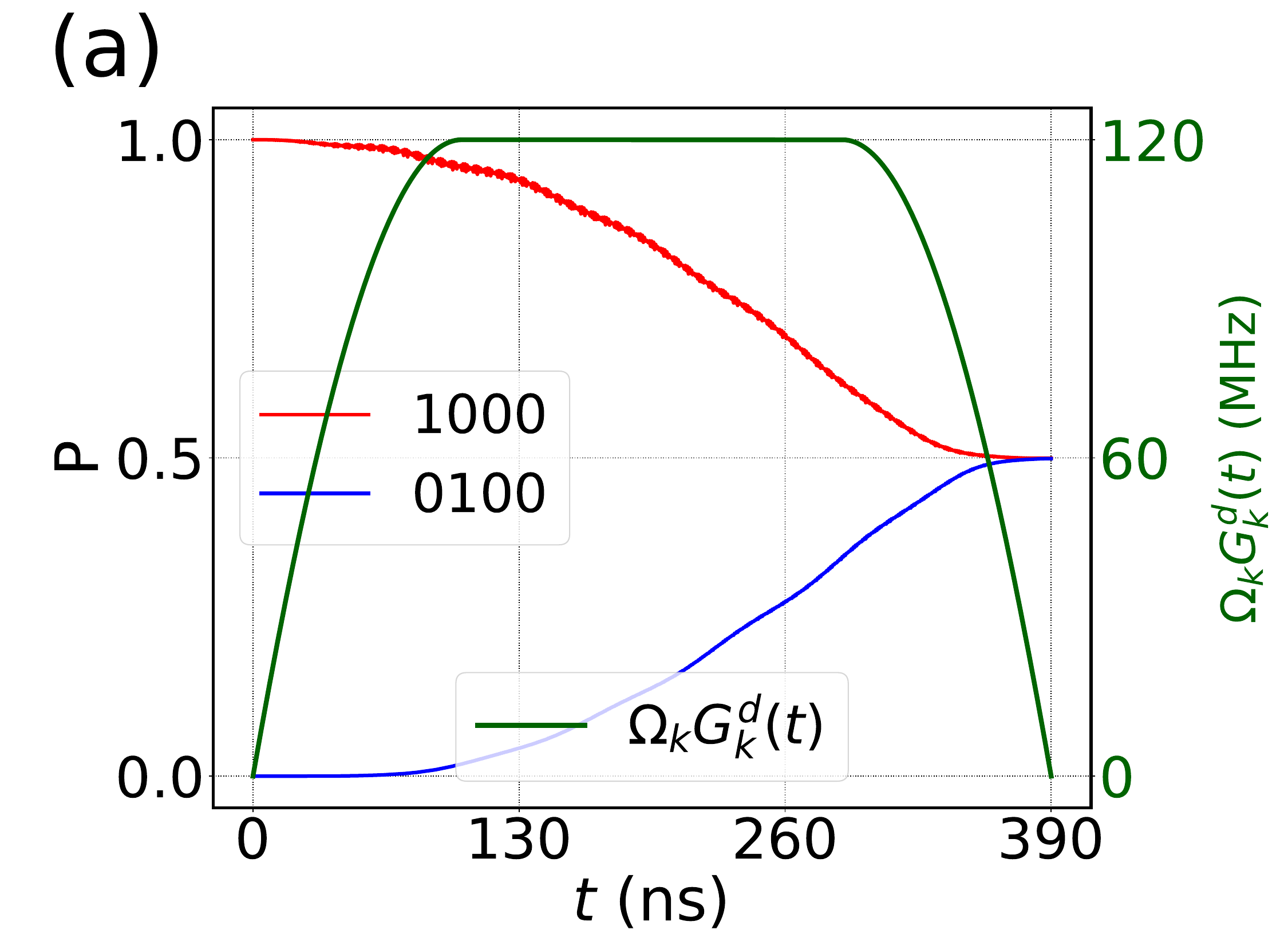}\includegraphics[width=4.0cm, height=4.05cm]{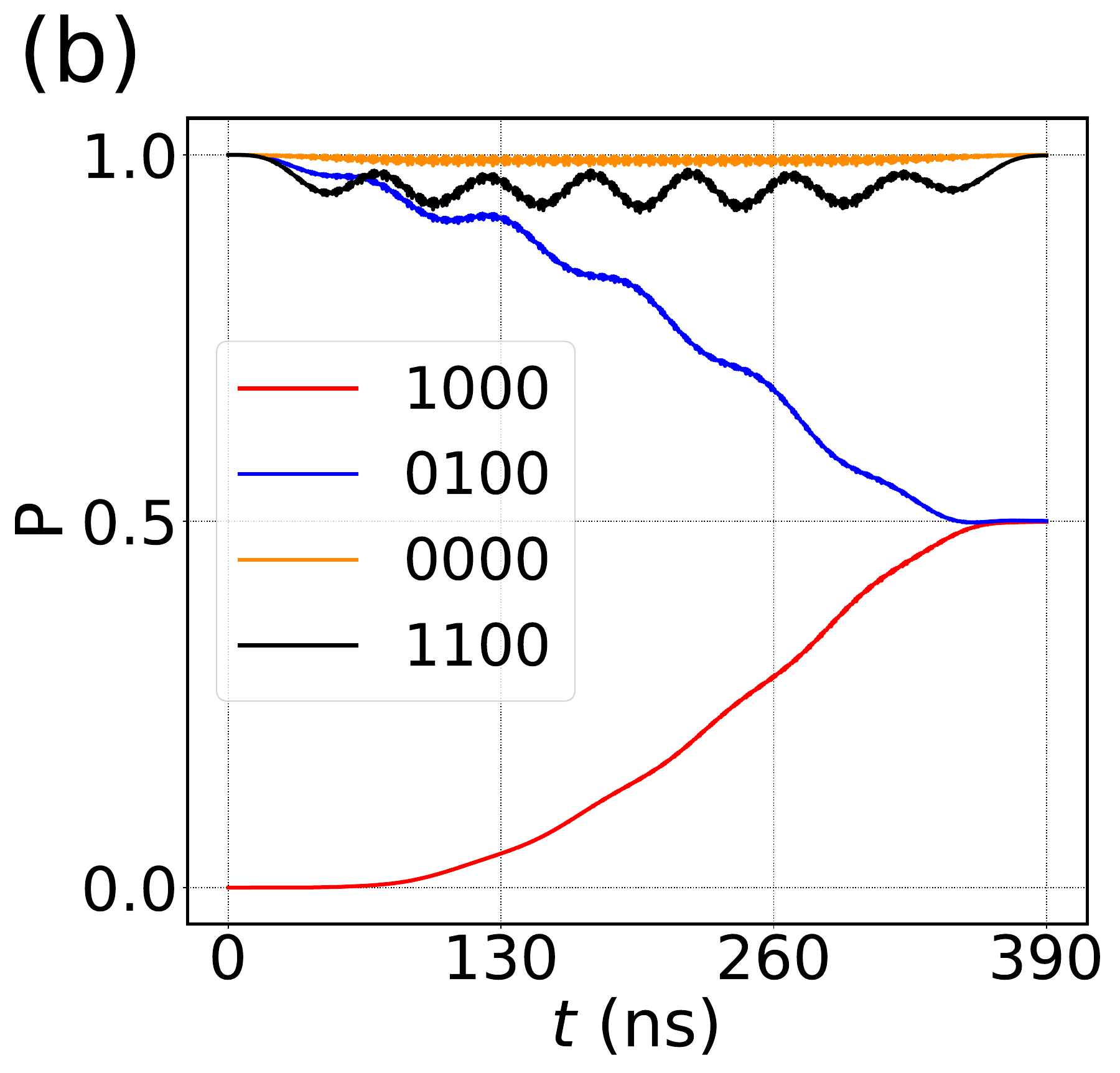}
\includegraphics[width=4.60cm, height=4.05cm]{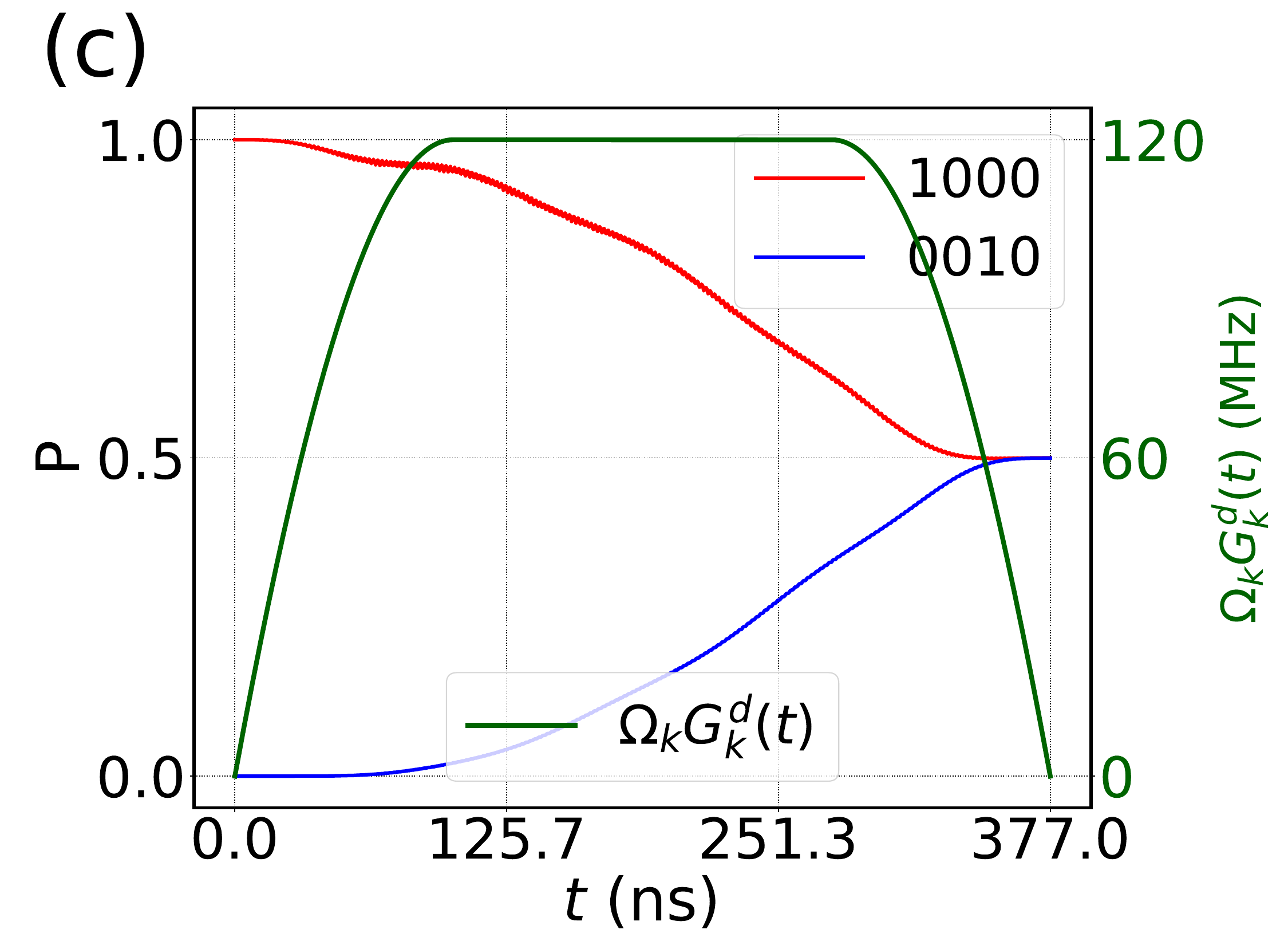}\includegraphics[width=4.0cm, height=4.05cm]{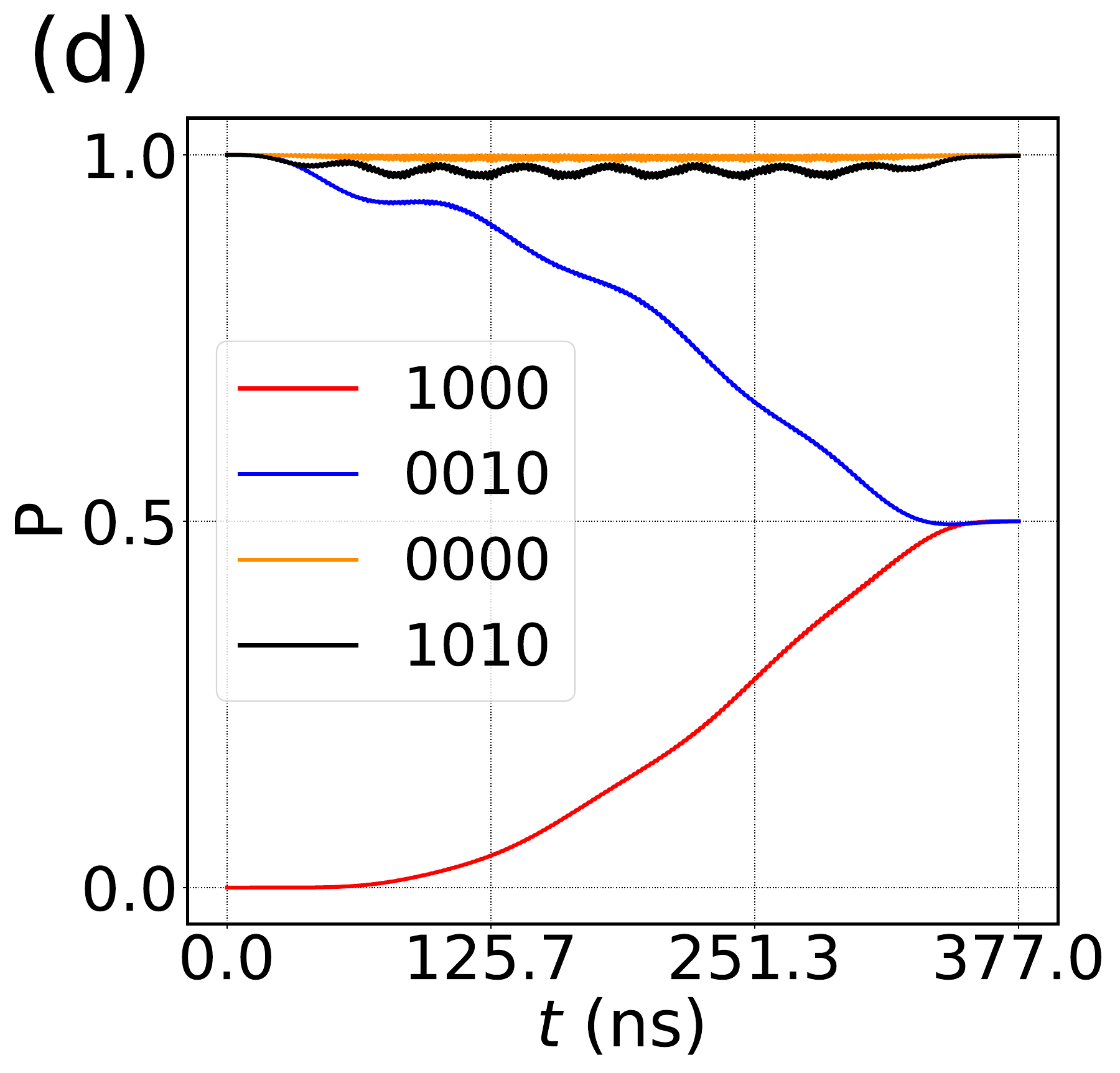}
\end{center}
\caption{The dynamics of the system during the gate operation for the three-qubit case. Populations of states $|1000\rangle$ (in red) and $|0100\rangle$ (in blue) are presented in (a) with the initial state $|1000\rangle$, while changes in populations of states $|1000\rangle$, $|0100\rangle$, $|0000\rangle$ (in orange), and $|1100\rangle$ (in black) are illustrated in (b), corresponding to the initial states $|0100\rangle$, $|0000\rangle$, and $|1100\rangle$, respectively. The green line represents the applied pulse shape with parameter values $\Omega_{1(2)}(0)/2\pi = 120$ MHz, $\omega^d_1/2\pi \approx 8.694$ GHz, $\omega^d_2/2\pi \approx 7.782$ GHz, $\gamma = 204$ ns, with a fixed relation $t_r = 0.5\gamma$. The gate operation between $Q_1$ and $Q_3$, along with the populations of states $|1000\rangle$ (in red), $|0010\rangle$ (in blue) and the applied time-dependent pulse shape (in green), are shown in (c), while changes in populations for states $|1000\rangle$, $|0010\rangle$, $|0000\rangle$ (in orange), and $|1010\rangle$ (in black) during the system dynamics are presented in (d) with pulse parameters set to $\Omega_{1(2)}(0)/2\pi = 120$ MHz, $\omega^d_1/2\pi \approx 8.701$ GHz, $\omega^d_2/2\pi \approx 8.194$ GHz, $\gamma = 202$ ns, and a fixed relation $t_r = 0.5\gamma$. The notation $|Q_1, Q_2, Q_3, Q_c\rangle$ indicates the states of $Q_1$, $Q_2$, $Q_3$, and $Q_c$, respectively. The system parameter values are provided in Table III.
\label{FigNine}}
\vspace{3.5 em}  
\end{figure}

\section{Analysis of $\sqrt{i\text{SWAP}}$ gate for a three qubits system}

Here, we explore the implementation of a $\sqrt{i\text{SWAP}}$ gate between arbitrary qubit pairs for a three-qubit system, in which three transmon qubits are coupled to a single transmon coupler. The system parameters are detailed in Table III. To achieve the $\sqrt{i\text{SWAP}}$ operation between qubits $Q_1$ and $Q_2$, the frequencies of the two drives applied to the coupler should be carefully chosen as $\omega^d_1/2\pi \approx 2\omega_c + \alpha_c - \omega_1 + \delta$ and $\omega^d_2/2\pi \approx 2\omega_c + \alpha_c - \omega_2 + \delta$. Following our discussions, one drive frequency is determined as $\omega^d_1 \approx 8.694$ GHz, with the chosen detuning $\delta/2\pi = -20$ MHz, drive amplitude $\Omega_{1(2)}(0)/2\pi = 120$ MHz, and system parameters from Table III. By investigating the dynamics of the system and sweeping another drive frequency, we obtain $\omega^d_2/2\pi \approx 7.782$ GHz. Employing the microwave pulse shape with the specified relation $t_r = 0.5\gamma$ and parameter values $\gamma = 204$ ns, we successfully execute the $\sqrt{i\text{SWAP}}$ gate operation between $Q_1$ and $Q_2$, as depicted in Figs. 9(a,b). Notably, maximal entanglement between $Q_1$ and $Q_2$ is achieved after the gate operation, while the states $|0000\rangle$ and $|1100\rangle$ (corresponding to the states of $Q_1$, $Q_2$, $Q_3$, and $Q_c$ successively) remain unchanged. Up to the single-qubit phase, the fidelity of $\sqrt{i\text{SWAP}}$ gate is only 68.12\%. However, considering both single-qubit phase compensation and introducing a conditional phase of $\phi$ $\approx$ 0.409 rad, the gate fidelity can reach as high as 99.93\%.

Furthermore, we explore the implementation of a $\sqrt{i\text{SWAP}}$ operation between $Q_1$ and $Q_3$. Based on the preceding analysis, we determine the microwave drive parameters necessary to achieve high-fidelity $\sqrt{i\text{SWAP}}$ gate operations between $Q_1$ and $Q_3$. Setting the detuning to $\delta/2\pi = -13$ MHz and utilizing the system parameters outlined in Table III, we estimate the drive frequency $\omega^d_1/2\pi \approx 8.701$ GHz. By scanning another drive frequency during system dynamics, we identify $\omega^d_2/2\pi \approx 8.194$ GHz. Employing pulse shape parameters $\Omega_{1(2)}(0)/2\pi = 120$ MHz, $t_r = 0.5 \gamma$, and $\gamma = 202$ ns, we illustrate the system dynamics during the $\sqrt{i\text{SWAP}}$ gate operation in Figs. 9(c-d). From Fig. 9(c), it is evident that maximal entanglement between $Q_1$ and $Q_3$ is achieved after the gate operation. As depicted in Fig. 9(d), the system state $|0000\rangle$ is significantly suppressed throughout the gate operation, while state $|1010\rangle$ exhibits slight oscillations due to coupling interactions with other states, ultimately returning to its initial state after the gate operation. Up to the single-qubit phases compensation, we ascertain that the average fidelity of the $\sqrt{i\text{SWAP}}$ gate between $Q_1$ and $Q_3$ reaches $F$ = 99.56\%. Introducing an additional conditional phase $\phi$ $\approx$ 0.281 rad boosts the gate fidelity to over 99.95\%.


\begin{table}[h]
\centering
\caption{System parameters for implementing a two-qubit $\sqrt{i\text{SWAP}}$ gate in a three-qubit system.}
\label{my-label}
\begin{tabular}{@{}lccccc@{}}
\toprule
              & Bare frequency (GHz) & Anharmonicity (MHz)   & Coupling (MHz)  \\
\hline
$Q_1$ &  $\omega_1/2\pi$ = 5.0         &          $\alpha_1/2\pi = -200$      &$g_{1c}/2\pi$ = 190    \\
$Q_2$     & $\omega_2/2\pi$ =  5.9     &          $\alpha_2/2\pi = -200$      &$g_{2c}/2\pi$ = 100  \\
$Q_3$     & $\omega_3/2\pi$ =  5.5     &          $\alpha_3/2\pi = -200$      &$g_{3c}/2\pi$ = 140  \\
$Q_c$ &  $\omega_c/2\pi$ = 7.0         &          $\alpha_c/2\pi = -400$      &                               \\  
\toprule
\end{tabular}
\end{table}


\section*{References}

\begin{thebibliography}{99}



\bibitem{ShorPW1994} P. W. Shor, Algorithms for quantum computation: discrete logarithms and factoring, In: Proceedings of the 35th Annual Symposium on Foundations of Computer Science, 124–134, (1994).
\bibitem{ShorPW1997} P. W. Shor, Polynomial-Time Algorithms for Prime Factorization and Discrete Logarithms on a Quantum Computer, SIAM J. Comp. {\bf26}, 1484 (1997).
\bibitem{WuYPRL2021} Y. Wu, W.-S.  Bao, S.  Cao, F. Chen, M. C. Chen, X. Chen, T. H. Chung, H. Deng, Y.  Du, D.  Fan, \emph{et al.,} Strong Quantum Computational Advantage Using a Superconducting Quantum Processor, Phys. Rev. Lett. {\bf127}, 180501 (2021).
\bibitem{Bermejo-VegaPRX2018} J.  Bermejo-Vega, D.  Hangleiter, M. Schwarz, R.  Raussendorf, J. Eisert, Architectures for Quantum Simulation Showing a Quantum Speedup, Phys. Rev. X {\bf8}, 021010 (2018).
\bibitem{HuangHYScience2022} H.-Y.  Huang, M.  Broughton, J.  Cotler, S. Chen, J.  Li, M.  Mohseni, H. Neven, R.  Babbush, R.  Kueng, J. Preskill,  \emph{et al}., Quantum advantage in learning from experiments, Science {\bf376}, 1182 (2022).





\bibitem{JLOBrienScience2007} J. L. O. Brien, Optical Quantum Computing, Science {\bf318}, 1567 (2007).
\bibitem{GJPrydeAPR2019} S. Slussarenko and G. J. Pryde, Photonic quantum information processing: A concise review, Appl. Phys. Rev. {\bf6}, 041303 (2019).




\bibitem{RBlattPR2008}  H. H$\ddot{\text{a}}$ffner, C. F. Roos, and R. Blatt, Quantum computing with trapped ions, Phys. Rep. {\bf469}, 155 (2008).




\bibitem{KrantzPAPR2019} P. Krantz, M. Kjaergaard, F. Yan, \emph{et al}., A quantum engineer’s guide to superconducting qubits, Appl Phys Rev, {\bf6}, 021318 (2019).
\bibitem{KjaergaardM2019} M. Kjaergaard, M. E. Schwartz,  \emph{et al}., Superconducting qubits: current state of play, Annu Rev Condensed Matter Phys, {\bf11}, 369–395 (2019).




\bibitem{AlexandreBlaisPRA2004} A. Blais, R.-S. Huang, A. Wallraff, S. M. Girvin, and R. J. Schoelkopf, Cavity quantum electrodynamics for superconducting electrical circuits: An architecture for quantum computation, Phys. Rev. A {\bf69}, 062320 (2004).
\bibitem{AlexandreBlaisPRA2007} A. Blais, J. Gambetta, A. Wallraff, D. I. Schuster, S. M. Girvin, M. H. Devoret, and R. J. Schoelkopf, Quantum-information processing with circuit quantum electrodynamics, Phys. Rev. A {\bf75}, 032329 (2007).

\bibitem{JClarkeNature2008} J. Clarke and F. K. Wilhelm, Superconducting quantum bits, Nature {\bf453}, 1031 (2008).

\bibitem{ABlaisRMP2021} A. Blais, A. L. Grimsmo, S. M. Girvin, and A. Wallraff, Circuit quantum electrodynamics, Rev. Mod. Phys. {\bf93}, 025005 (2021).



\bibitem{JRPettaRMP2023} G. Burkard, T. D. Ladd, A. Pan, J. M. Nichol, and J. R. Petta, Semiconductor spin qubits, Rev. Mod. Phys. {\bf95}, 025003 (2023).







\bibitem{MASillanpNature2007} M. A. Sillanp$\ddot{\text{a}}\ddot{\text{a}}$, J. I. Park, and R. W. Simmonds, Coherent quantum state storage and transfer between two phase qubits via a resonant cavity, Nature {\bf449}, 438 (2007).
\bibitem{JMajerNature2007} J. Majer, J. M. Chow, J. M. Gambetta, J. Koch, \emph{et al}., Coupling superconducting qubits via a cavity bus, Nature {\bf449}, 443 (2007).
\bibitem{AONiskanenScience2007} A. O. Niskanen, K. Harrabi, \emph{et al}., Quantum Coherent Tunable Coupling of Superconducting Qubits, Science {\bf316}, 723 (2007).
\bibitem{PlantenbergNature2007} J. H. Plantenberg, P. C. de Groot, C. J. Harmans, and J. E. Mooij, Demonstration of controlled-NOT quantum gates on a pair of superconducting quantum bits, Nature {\bf447}, 836 (2007).
\bibitem{LDiCarloNature2009} L. DiCarlo, J. M. Chow, \emph{et al}., Demonstration of two-qubit algorithms with a superconducting quantum processor, Nature {\bf460}, 240 (2009).
\bibitem{MAnsmannNature2009} M. Ansmann, H. Wang, \emph{et al}., Violation of Bell's inequality in Josephson phase qubits, Nature {\bf461}, 504 (2009).



\bibitem{FAruteNature2019} F. Arute \emph{et al}., Quantum supremacy using a programmable superconducting processor, Nature {\bf574}, 505 (2019).



\bibitem{MGongScience2021} M. Gong \emph{et al}., Quantum walks on a programmable two- dimensional 62-qubit superconducting processor, Science {\bf372}, 948 (2021).


\bibitem{DRWYostnpjQI2020} D. R. W. Yost, M. E. Schwartz, \emph{et al}., Solid-state qubits integrated with superconducting through-silicon vias, npj Quant. Inf. {\bf6}, 59 (2020).
\bibitem{JLMallekarXiv2103} J. L. Mallek, D.-R. W. Yost, \emph{et al}., Fabrication of superconducting through-silicon vias, arXiv:2103.08536.
\bibitem{SKosenQST2022} S. Kosen \emph{et al}., blocks of a flip-chip integrated superconducting quantum processor. Quantum Sci. Technol. {\bf7}, 035018 (2022).




\bibitem{BoixoSNP2018} S.  Boixo, S. V.  Isakov, V. N.  Smelyanskiy, \emph{et al}., Characterizing quantum supremacy in near-term devices. Nat Phys, {\bf14}, 595–600 (2018).
\bibitem{PreskillJQauntum2018} J. Preskill, Quantum computing in the NISQ era and beyond. Quantum, {\bf2}, 79 (2018).



\bibitem{DDeutschPNAS1995} D. Deutsch, A. Barenco, and A. Ekert, Universality in quantum computation, Proc. R. Soc. Lond. A {\bf449}, 669 (1995).
\bibitem{ABarencoPNAS1995} A. Barenco, A universal two-bit gate for quantum computation, Proc. R. Soc. Lond. A {\bf449}, 679 (1995).
\bibitem{DPDiVincenzoPRA10151995} D. P. DiVincenzo, Two-bit gates are universal for quantum computation, Phys. Rev. A {\bf51}, 1015 (1995).
\bibitem{TSleatorPRL1995} T. Sleator and H. Weinfurter, Realizable Universal Quantum Logic Gates, Phys. Rev. Lett. {\bf74}, 4087 (1995).
\bibitem{SLloyd1995} S. Lloyd, Almost Any Quantum Logic Gate is Universal, Phys. Rev. Lett. {\bf75}, 346 (1995).
\bibitem{ABarencoPRA34571995} A. Barenco, C. H. Bennett, R. Cleve, D. P. DiVincenzo, N. Margolus, P. Shor, T. Sleator, J. A.Smolin, and H. Weinfurter, Elementary gates for quantum computation, Phys. Rev. A {\bf52}, 3457 (1995).


\bibitem{XGuPhysRep2017} X. Gu, A. F. Kockum, A. Miranowicz, Y.-x. Liu, and F. Nori, Microwave photonics with superconducting quantum circuits, Phys. Rep. {\bf718-719}, 1 (2017).





\bibitem{LDiCarloNature2010} L. DiCarlo, M. D. Reed, \emph{et al}., Preparation and measurement of three-qubit entanglement in a superconducting circuit, Nature {\bf467}, 574 (2010).
\bibitem{RBarendsNature2014} R. Barends, J. Kelly, \emph{et al}., Superconducting quantum circuits at the surface code threshold for fault tolerance, Nature {\bf508}, 500 (2014).


\bibitem{DCMcKayPRAppl2016} D. C. McKay, S. Filipp, \emph{et al}., Universal Gate for Fixed-Frequency Qubits Via a Tunable Bus, Phys. Rev. Appl. {\bf6}, 064007 (2016).
\bibitem{MRothPRA2017} M. Roth, M. Ganzhorn, N. Moll, S. Filipp, G. Salis, and S. Schmidt, Analysis of a parametrically driven exchange-type gate and a two-photon excitation gate between superconducting qubits, Phys. Rev. A {\bf96}, 062323 (2017).


\bibitem{YChenPRL2014} Y. Chen, C. Neill, P. Roushan, \emph{et al}., Qubit Architecture with High Coherence and Fast Tunable Coupling, Phys. Rev. Lett. {\bf113}, 220502 (2014).
\bibitem{FeiYanPRAp2018} F. Yan, P. Krantz, Y. Sung, M. Kjaergaard, D. L. Campbell, T. P. Orlando, S. Gustavsson, and W. D. Oliver, Tunable Coupling Scheme for Implementing High-Fidelity Two-Qubit Gates, Phys. Rev. Appl. {\bf10}, 054062 (2018).


\bibitem{JMChowNJP2013} J. M. Chow, J. M. Gambetta, A. W. Cross, S. T. Merkel, C. Rigetti, and M. Steffen, Microwave-activated conditional-phase gate for superconducting qubits, New J. Phys. {\bf15}, 115012 (2013).
\bibitem{EBarnesPRB2017} E. Barnes, C. Arenz, A. Pitchford, and S. E. Economou, Fast microwave-driven three-qubit gates for cavity-coupled superconducting qubits, Phys. Rev. B {\bf96}, 024504 (2017).
\bibitem{SPPremaratnePRA2019} S. P. Premaratne, J.-H. Yeh, F. C. Wellstood, and B. S. Palmer, Implementation of a generalized controlled-NOT gate between fixed-frequency transmons, Phys. Rev. A {\bf99}, 012317 (2019).

\bibitem{CRigettiPRB2010} C. Rigetti and M. Devoret, Fully microwave-tunable universal gates in superconducting qubits with linear couplings and fixed transition frequencies, Phys. Rev. B {\bf81}, 134507 (2010).
\bibitem{JMChowPRL2011} J. M. Chow, \emph{et al}., Simple All- Microwave Entangling Gate for Fixed-Frequency Superconducting Qubits, Phys. Rev. Lett. {\bf107}, 080502 (2011).
\bibitem{JLAllenPRA2017}  J. L. Allen, R. Kosut, J. Joo, P. Leek, and E. Ginossar, Optimal control of two qubits via a single cavity drive in circuit quantum electrodynamics, Phys. Rev. A {\bf95}, 042325 (2017). 
\bibitem{NakamuraPRL2606012023} S. Shirai, Y. Okubo, K. Matsuura, A. Osada, Y. Nakamura, and A. Noguchi, All-microwave manipulation of superconducting qubits with a fixed-frequency transmon coupler, Phys. Rev. Lett. {\bf130}, 260601 (2023).



\bibitem{JianxinChenPRL130070601} C. Huang, T. Wang, F. Wu, D. Ding, \emph{et. al}., Quantum Instruction Set Design for Performance, Phys. Rev. Lett. {\bf130}, 070601 (2023).


\bibitem{BarendsPRL2105012019} R. Barends \emph{et al}, Diabatic gates for frequency-tunable superconducting qubits, Phys. Rev. Lett. {\bf123}, 210501 (2019).

\bibitem{BFoxenPRL125120504} B. Foxen \emph{et al}., (Google AI Quantum Collaboration), Demonstrating a Continuous Set of Two-Qubit Gates for Near-Term Quantum Algorithms, Phys. Rev. Lett. {\bf125}, 120504 (2020).

\bibitem{MGVavilovPRXQ2020345} K. N. Nesterov, Q. Ficheux, V. E. Manucharyan, and M. G. Vavilov, Proposal for Entangling Gates on Fluxonium Qubits via a Two-Photon Transition, PRX Quantum {\bf2}, 020345 (2021).

\bibitem{MJSPRA108052619} J. J. C\'{a}ceres, D. Dom\'{i}nguez, and M. J. S\'{a}nchez, Fast quantum gates based on Landau-Zener-St$\ddot{\text{u}}$ckelberg-Majorana transitions, Phys. Rev. A {\bf108}, 052619, (2023).




\bibitem{JKochPRA0423192007} J. Koch, T. M. Yu, \emph{et al}., Charge-insensitive qubit design derived from the Cooper pair box, Phys. Rev. A {\bf76}, 042319 (2007).



\bibitem{KlausAndersPRA0223112000}  A. S{\o}rensen and K. M{\o}lmer, Entanglement and quantum computation with ions in thermal motion, Phys. Rev. A {\bf62}, 022311 (2000).




\bibitem{AFKockumPRA0638492017} A. F. Kockum, A. Miranowicz, \emph{et al}.,  Deterministic quantum nonlinear optics with single atoms and virtual photons, Phys. Rev. A {\bf95}, 063849 (2017).







\bibitem{PengZhaoAPL2022} P Zhao, Y Zhang, G Xue, Y Jin, H Yu, Tunable coupling of widely separated superconducting qubits: A possible application toward a modular quantum device, Appl. Phys. Lett. {\bf121}, 032601 (2022).
\bibitem{ChenYinqiPRAp2022} Y. Chen, K. N. Nesterov, V. E. Manucharyan, and M. G. Vavilov, Fast Flux Entangling Gate for Fluxonium Circuits, Phys. Rev. Applied {\bf18}, 034027 (2022).



\bibitem{YuHaifeng2022} C. Wang, X. Li, H. Xu, \emph{et al}., Towards practical quantum computers: transmon qubit with a lifetime approaching 0.5 milliseconds. npj Quantum Inf {\bf8}, 3 (2022).

\bibitem{PlaceNC2021} A. P. M. Place, L. V. H. Rodgers, P. Mundada, B. M. Smitham, M. Fitzpatrick, Z. Leng, A. Premkumar, J. Bryon, S. Sussman, G. Cheng, \emph{et al}., New material platform for superconducting transmon qubits with coherence times exceeding 0.3 milliseconds, Nat. Commun. {\bf12}, 1779 (2021).



\bibitem{FeiYanPRL129040502} Z. Ni, S. Li, L. Zhang, J. Chu, J. Niu, T. Yan, X. Deng, L. Hu, J. Li, Y. Zhong, S. Liu, F. Yan, Y. Xu, and D. Yu, Scalable Method for Eliminating Residual ZZ Interaction between Superconducting Qubits, Phys. Rev. Lett. {\bf129}, 040502 (2022).


\bibitem{ZhaoPengPRA952017} P. Zhao, X. Tan, H. Yu, S.-L. Zhu, and Y. Yu, Simultaneously exciting two atoms with photon-mediated Raman interactions, Phys. Rev. A {\bf95}, 063848 (2017).
\bibitem{PengZhaoPRA960438332017} P. Zhao, X. Tan, H. Yu, S.-L. Zhu, and Y. Yu, Circuit QED with qutrits: Coupling three or more atoms via virtual-photon exchange, Phys. Rev. A {\bf96}, 043833 (2017).



\bibitem{XuPengPRA1082023} P. Xu, Q. Jing, P. Zhao, and Y. Yu, Microwave-driven iSWAP-like gate for fixed-frequency superconducting transmon qutrits, Phys. Rev. A {\bf108}, 032615 (2023).





\bibitem{PedersenPLA2007} L. H. Pedersen, N. M. M{\o}ller, and K. M{\o}lmer, Fidelity of quantum operations, Phys. Lett. A {\bf367}, 47 (2007).



\bibitem{GhoshPRA2013} J. Ghosh, A. Galiautdinov, Z. Zhou, A. N. Korotkov, J. M. Martinis and M. R. Geller, High-fidelity controlled-$\sigma^z$ gate for resonator-based superconducting quantum computers, Phys. Rev. A {\bf87}, 022309 (2013).



\bibitem{ZahedinejadPRAp2016} E. Zahedinejad, J. Ghosh, and B. C. Sanders, Designing High-Fidelity Single-Shot Three-Qubit Gates: A Machine-Learning Approach, Phys. Rev. Appl. {\bf6}, 054005 (2016).
\bibitem{BarnesPRB2017} E. Barnes, C. Arenz, A. Pitchford, and S. E. Economou, Fast microwave-driven three-qubit gates for cavity-coupled superconducting qubits, Phys. Rev. B {\bf96}, 024504 (2017).




\bibitem{JayMGambettaPRA970323062018} C. J. Wood and J. M. Gambetta, Quantification and characterization of leakage errors, Phys. Rev. A {\bf97}, 032306 (2018).



\bibitem{JHShirleyB9791965} J. H. Shirley, Solution of the Schr$\ddot{\text{o}}$dinger Equation with a Hamiltonian Periodic in Time, Phys. Rev. {\bf138}, B979 (1965).
\bibitem{YBZelDovichJETP10061967} Y. B. Zel'Dovich, The quasienergy of a quantum-mechanical system subjected to a periodic action, Sov. Phys. JETP {\bf24}, 1006 (1967).
\bibitem{VIRitusJETP1967} V. I. Ritus, Shift and splitting of atomic energy levels by the field of an electromagnetic wave, Sov. Phys. JETP {\bf24}, 1041 (1967).
\bibitem{HSambePRA722031973} H. Sambe, Steady States and Quasienergies of a Quantum-Mechanical System in an Oscillating Field, Phys. Rev. A {\bf7}, 2203 (1973).





\end{thebibliography}
\end{document}